\documentclass[12pt]{article}

\usepackage{amsmath, amsthm, amsfonts, bm}
\usepackage{amssymb}
\usepackage{mathtools}

\usepackage{graphicx,psfrag,epsf, float}
\usepackage{enumerate,titlesec, color}
\usepackage{natbib}
\usepackage{footnote}
\usepackage{url} 
\usepackage{soul}
\usepackage{tikz}
\usepackage[ruled,linesnumbered]{algorithm2e}
\usepackage[font=small,labelfont=bf]{caption}

\usepackage{float}
\usepackage{lineno}
\usepackage{soul}

\usepackage{setspace}

\usepackage{geometry}

\renewcommand\footnotemark{}

\usepackage{booktabs}

\usepackage{array}

\usepackage{multirow}
\definecolor{napiergreen}{rgb}{0.16, 0.5, 0.0}
\definecolor{myrtle}{rgb}{0.13, 0.26, 0.12}
\definecolor{dartmouthgreen}{rgb}{0.05, 0.5, 0.06}

\newtheorem{prop}{Proposition}

\def\bfa{\mathbf a}

\def\bfd{\mathbf d}

\def\bff{\mathbf f}
\def\bfg{\mathbf g}
\def\bfh{\mathbf h}
\def\bfx{\mathbf x}
\def\bfy{\mathbf y}
\def\bfz{\mathbf z}
\def\bfs{\mathbf s}

\def\bfA{\mathbf A}
\def\bfB{\mathbf B}

\def\bfX{\mathbf X}

\def\bfU{\mathbf U}

\def\bfu{\mathbf u}

\def\bfalpha{\boldsymbol \alpha}
\def\bfbeta{\boldsymbol \beta}

\def\bfepsilon{\boldsymbol \epsilon}

\def\bfPhi{\boldsymbol \Phi}

\def\bfTheta{\boldsymbol \Theta}

\def\bflambda{\boldsymbol\lambda}
\def\bfLambda{\boldsymbol\Lambda}

\def\bfomega{\boldsymbol\omega}

\def\bfpsi{\boldsymbol\psi}
\def\real{\mathbb R}

\def\bfzero{\boldsymbol 0}
\def\bfone{\boldsymbol 1}

\def\cA{\mathcal A}
\def\cB{\mathcal B}

\def\cG{\mathcal G}
\def\mbE{\mathbb E}

\def\cB{\mathcal B}

\def\cG{\mathcal G}

\def\diag{\mbox{diag}}

\def\prox{\mbox{prox}}
\def\argmin{\mbox{argmin}}

\begin{document}

	\bigskip
	\date{}
	\title{Genetic underpinnings of brain structural connectome for young adults}

	\author{{Yize Zhao$^1$$^*$},
    {Changgee Chang$^2$},
    {Jingwen Zhang$^3$},  
    {and Zhengwu Zhang$^4$$^*$}\bigskip\\
   $^1$\small{Department of Biostatistics, Yale University}\\
$^2$\small{Department of Biostatistics, Epidemiology and Informatics,} \small{ University of Pennsylvania}\\
$^3$\small{Department of Biostatistics,  Boston University, Boston, MA.}\\
    $^4$\small{Department of Statistics and Operations Research,} \small{University of North Carolina at Chapel Hill}\\
   \footnote{*Correspondence should be directed to: yize.zhao@yale.edu and zhengwu\_zhang@unc.edu}}

	\maketitle

\def\spacingset#1{\renewcommand{\baselinestretch}%
	{#1}\small\normalsize} \spacingset{1}

\vspace{-0.5in}
\begin{abstract} 
With distinct advantages in power over behavioral phenotypes, brain imaging traits have become emerging endophenotypes to dissect molecular contributions to behaviors and neuropsychiatric illnesses. Among different imaging features, brain structural connectivity (i.e., structural connectome) which summarizes the anatomical connections between different brain regions  is one of the most cutting edge while under-investigated traits; and the genetic influence on the structural connectome variation remains highly elusive.  Relying on a landmark imaging genetics study for young adults, we develop a biologically plausible brain network response shrinkage model to comprehensively characterize the relationship between high dimensional genetic variants and the structural connectome phenotype. Under a unified Bayesian framework, we accommodate the topology of brain network and biological architecture within the genome; and eventually establish a mechanistic mapping between genetic biomarkers and the associated brain sub-network units. An efficient  expectation-maximization algorithm is developed to estimate the model and ensure computing feasibility. In the application to  the Human Connectome Project 
Young Adult  (HCP-YA) data, we establish the genetic underpinnings which are highly interpretable under functional annotation and brain tissue eQTL analysis, for the brain white matter tracts  connecting the hippocampus and two cerebral hemispheres. We also show the superiority of our method in extensive simulations.   
\end{abstract}
\vspace{1cm}
\noindent%
\newpage
\spacingset{1.2}

\section{Introduction}
Brain imaging genetics, which studies the relationship between genetic alternation and brain structural or functional variation, has become an emerging research field in recent years \citep{elliott2018genome,nathoo2019review,shen2019brain}. Compared with traditional behavioral outcomes,  imaging quantitative traits  have a distinct advantage in capturing neurological etiology and have yielded prominent new findings in studying both psychiatric disorders \citep{ramanan2014apoe} and healthy populations \citep{zhao2019large}.

\begin{figure}[h]
\centering

\includegraphics[height=6cm]{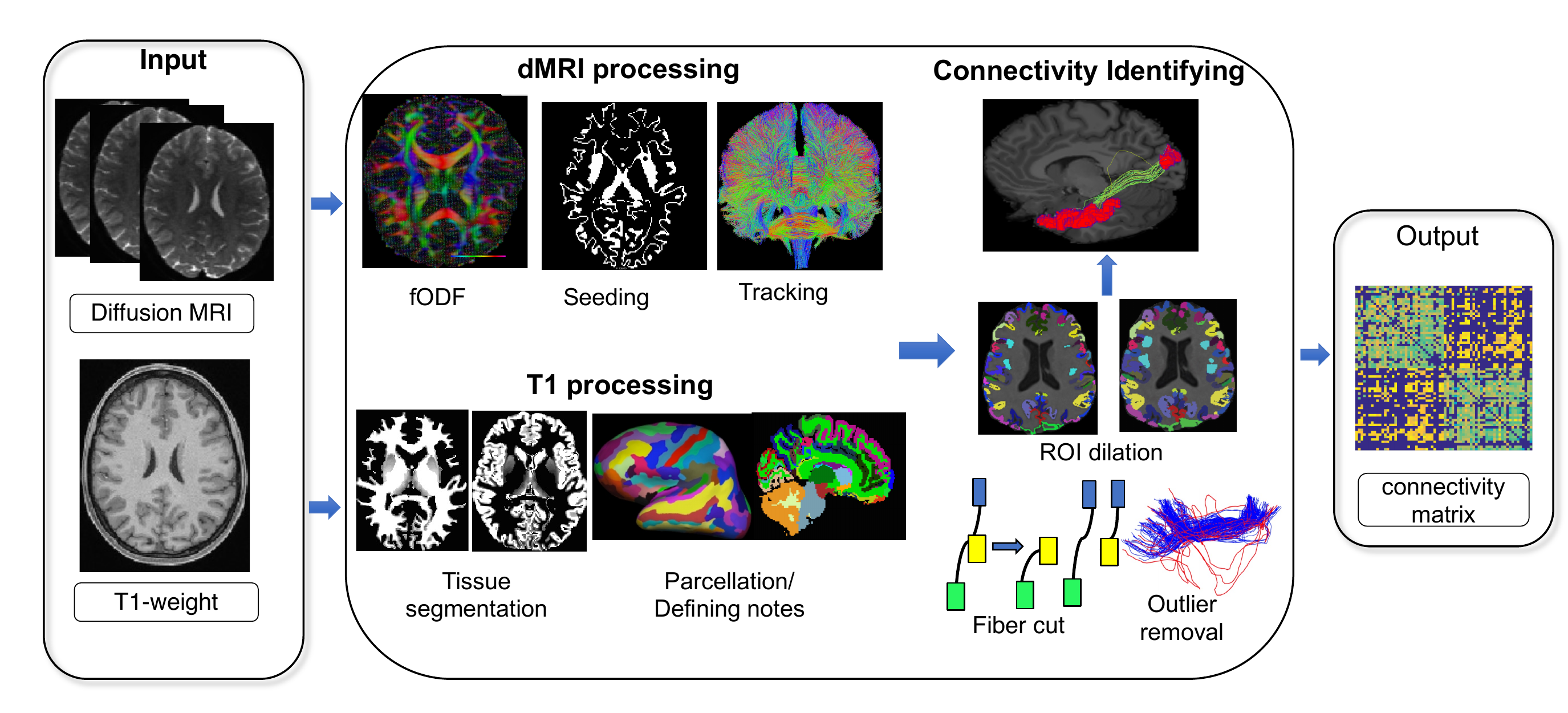}
\vspace{-0.5cm}
    \caption{s{Estimating brain structural connectivity from  diffusion MRI and T1}.  }
    \label{fig:image_processing}
\end{figure}

Among different types of  brain imaging characteristics, twin studies have found that brain structural features are highly heritable \citep{lenroot2008changing}.
Lately, using classical genome-wide association studies (GWAS), a few papers have revealed genetic predisposition for cortical and subcortical volume and shape variation \citep{elliott2018genome},  and have identified a number of associated loci.  Compared to traditional structural imaging phenotypes, brain structural connectivity, i.e., the network formed by white matter fiber tracts across the brain \citep{park2013structural}, is one of the most cutting edge but under-investigated features. Estimated from diffusion-weighted magnetic resonance imaging (dMRI) data, structural connectivity aims to measure the extent to which brain regions are interconnected by white matter fiber bundles and has been adopted to understand  its associations with cognition \citep{WANG2021117493}, aging \citep{antonenko2013functional}, and mental disorders \citep{zhang2016neural}. {To demonstrate, Figure \ref{fig:image_processing} displays the process of extracting the structural connectivity or network from dMRI data. The steps include estimating fiber orientation distribution function (fODF) for each voxel, segmenting the brain into regions of interest (ROIs), performing fiber tracking, and identifying connections between ROIs. More details on this process can be found in Section \ref{sec:data}. In this paper, we are interested in studying how genetics influence an individual's structural network  by simultaneously dissecting the connectome-related genetic underpinnings and the corresponding brain sub-network signatures.}

It is crucial yet challenging to model brain networks properly. In both clinical and neuroscience fields, one primary strategy to handle network data is to independently analyze the individual connections between each pair of nodes \citep{shen2017using}. Such approaches completely ignore the topographic information within the network, leading to a loss of analytical power. Within the statistical community, some recent works realize the importance of considering the network topology and have started to handle networks using various combinations of low-rank and sparse models. For example, \cite{durante2018bayesian} construct a nonparametric Bayesian framework where the networks are represented through a mixture of low-rank factorizations; \cite{zhang2018network} and \cite{zhao2019covariate} explicitly consider a brain network as the response under a regression or generalized regression model and adopt low rank and/or sparsity constraints to estimate the coefficients. \cite{hu2019nonparametric} further extend the model to a nonparametric framework and perform a group comparison of dynamic networks. While the above methods provide decent solutions to quantify brain network variation, they can only accommodate low dimensional covariates.  

Currently, the genetic associations of variations in the brain structural connectome remain highly elusive, which impedes our understanding of the molecular mechanisms supporting brain structural signatures during brain development and psychiatric illness onset.
Previous investigations on this topic are limited and heavily concentrated on summarized network features \citep{giddaluru2016genetics,elsheikh2020genome}. When we expand our view to studies that tackle any type of brain networks,  we find several recent works which associate brain functional connectivity with genetic variants \citep{richiardi2015correlated,li2020super}. Among those studies, the majority are based on a massive univariate analysis by marginally associating each single nucleotide polymorphism (SNP) with each network edge. The drawbacks of such a strategy are obvious -- besides overlooking the topographic information within the brain network and genetic correlation within the genome, this leads to a prohibitive burden of multiple comparison and computation with a massive number of univariate regressions. Beyond the univariate analysis, attempts have been made to infer the joint effect from either the genotype or phenotype side \citep{li2020super,tao2017generalized}. The Low-Rank Linear Regression Model (L2RM) by \cite{kong2020l2rm} is the only work we know of that jointly models a brain network matrix and high dimensional genetic features.  L2RM consists of a step-wise procedure with an initial screening to  downsize candidate SNPs followed by a low rank regularization. In this work,  however, we perform a unified analysis under a Bayesian framework to directly integrate large-scale genetic variants and a high-dimensional network response. More importantly, the proposed estimation scheme offers a clear biological interpretation, enhancing its robustness and translational impact.

Specifically, we propose a Bayesian generalized network response regression to associate high-dimensional SNPs with the brain structural connectome.  To uncover genetic biomarkers and identify network endophenotypes, we develop a biological architecture-driven network shrinkage procedure to identify genetic factors and their associated brain sub-networks simultaneously. Here, the sub-networks are assumed to be  clique sub-graphs \citep{WANG2021117493}. This aligns with the biological insight that each biomarker is associated with a set of interrelated brain regions to provide the most efficient neural support for function and behavior \citep{bassett2017network}. To further improve the statistical power from potentially diluted signals and enhance biological plausibility, we also incorporate the population-level brain connectivity and the SNP-set information from biological blocks. Integrating the biological information from both data sources, we accomplish knowledge-driven genotype and network-phenotype selections in a unified Bayesian paradigm.

Our contributions are multi-fold. First, we are among the very first to
explore brain structural connectome endophenotypes and dissect their associated genetic underpinnings from studying healthy individuals.  
Unlike previous related studies, we simultaneously accommodate a network-valued phenotype and high-dimensional structured genotypes in a unified Bayesian framework. We process more than $1000$ subjects' brain imaging and genetic data from the Human Connectome Project Young Adult (HCP-YA) dataset \citep{van2013wu} and apply the proposed method to uncover genetic influence on the brain structural network. Second, we propose an innovative network response structural shrinkage (NRSS) prior to incorporate known biological information from both imaging and genetic data. On the genetic side, NRSS is constructed by splitting shrinkage effects into those at the SNP-set level, brain sub-network level and SNP-to-network level. On the brain network side, NRSS is able to incorporate the prior information from the population-level connectivity to enhance biological plausibility of the inferred SNP-network effect. We show in our simulations that the proposed model dramatically improves the estimation and feature selection accuracy. Moreover, the identified genetic signals and brain sub-networks from HCP-YA reveal a remarkable consistency with the scientific literature. Finally, to improve computational efficiency, we develop an efficient expectation-maximization (EM) algorithm to perform posterior inference. Our experiments show the proposed algorithm can significantly reduce the computational burden, 
ensuring a broad use of the proposed method in mental health and neurological and psychiatric studies. 

The remainder of the paper is organized as follows. In Section \ref{sec:data} we describe our imaging and genetics features from the HCP-YA. In Section \ref{sec:method}, we introduce the model formulation (\ref{sec:model}), propose the new prior (\ref{sec:NRSS}) with operating properties (\ref{sec:operating}), and develop an EM algorithm (\ref{sec:EM}). {We perform the HCP-YA data analysis in Section \ref{sec:realdataHCP-YA} followed by simulations in Section \ref{sec:simulation}}. Finally, we conclude in Section \ref{sec:dicussion}.

\section{Human Connectome Project Young Adult Data} \label{sec:data}

 The HCP-YA aims to characterize brain connectivity in young adults and enables detailed comparisons between brain circuits, behaviors, and genetics at a subject level \citep{VanEssen20122222}.  The prepossessed imaging data are accessed through the ConnectomeDB, and the genetic data are hosted at the database of Genotypes and Phenotypes (dbGaP).

\subsection{Brain structural connectome extraction} 
{We follow the procedure in Figure \ref{fig:image_processing} to extract the brain structural connectome. For each subject in HCP-YA, we download their dMRI and T1 data.} A full dMRI session includes 6 runs,
using 3 different gradient tables, with each table acquired once with right-to-left and left-to-right phase encoding polarities, respectively. Each gradient table includes approximately $90$
diffusion weighting directions plus 6 $b_0$ acquisitions interspersed throughout each
run.  Within each run, there are  three shells of $b=1000, 2000$, and $3000$ s/mm$^2$ interspersed
with an approximately equal number of acquisitions on each shell. The scans were done by using the spin echo EPI sequence on a 3T customized Connectome Scanner. Such settings give the final acquired image an isotropic voxel size of $1.25$ mm, and $270$ diffusion weighted scans distributed equally over $3$ shells. The T1 image has 0.7 mm$^3$ isotropic resolution. See \cite{VanEssen20122222} for more details about the data acquisition and the minimal preprocessing pipeline. {For more details on the dMRI and structural connectome, we refer interested readers to the introductory article by \cite{webster2015high}.}

\begin{figure}[h]
\centering

\includegraphics[height=4cm]{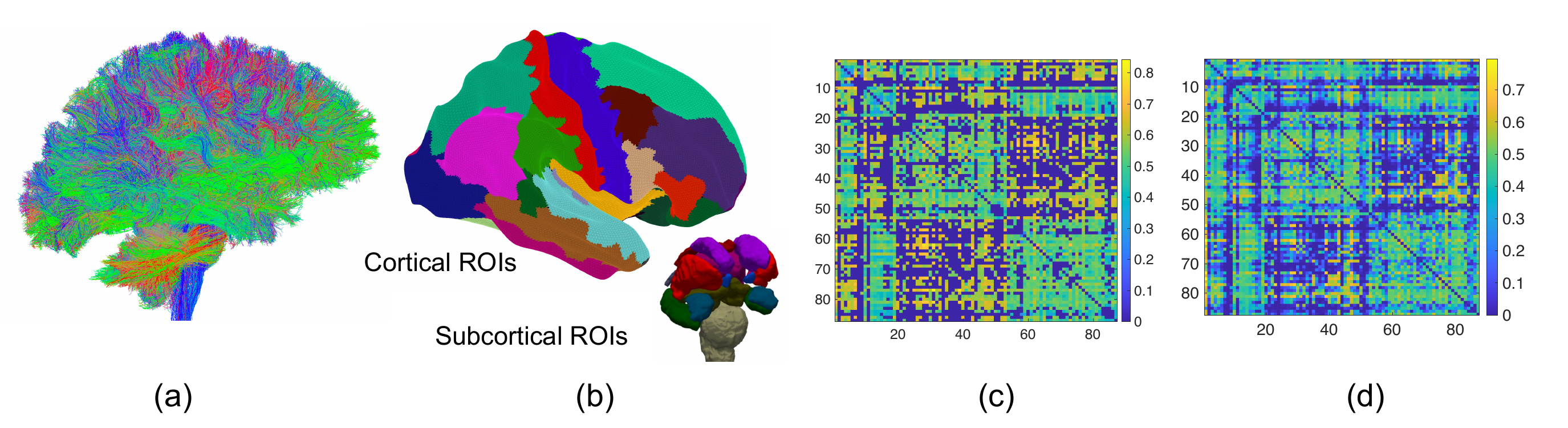}
\vspace{-0.2cm}
    \caption{An example of tractography result in (a),  DK parcellation  in (b), and   FA connectivity matrix  according to the DK parcellation in (c) from a single HCP-YA subject. Panel (d) shows the population mean connectivity of all HCP-YA subjects.}
    \label{fig:streamline}
\end{figure}

We apply a recent connectome mapping framework, named population-based structural connectome mapping (PSC) \citep{Zhang2017HCP}, to the minimally prepossessed dMRI and T1 data to extract the structural connectome for each subject. PSC uses a reproducible probabilistic tractography algorithm
\citep{maier2016tractography} to generate  whole-brain tractography data, which borrows anatomical information from high-resolution T1 image to reduce bias in the tractography. We use Desikan-Killiany (DK) atlas \citep{Desikan2006968} to define the ROIs corresponding to the nodes in the structural connectome. The DK parcellation contains 68 cortical surface regions with 34 nodes in each hemisphere, and 19 subcortical regions. The full ROI list is provided in the supplementary materials. For each pair of ROIs,  we extract the streamlines connecting them using the following procedure: 1) each gray matter ROI is dilated to include a small portion of white matter regions, 2) streamlines connecting multiple ROIs are cut into pieces so that we could extract the correct and complete pathway and 3) apparent outlier streamlines are removed.   {Given its wide application in brain imaging-genetic studies \citep{chiang2011genetics,zhao2021common}, mean fractional anisotropy (FA) value along streamlines is used to quantify connectivity strength in this paper.} {Figure \ref{fig:streamline} uses one HCP-YA subject to illustrate the tractography result (streamlines in panel (a)), the DK parcellation (in panel (b)),  and the final mean FA connectivity matrix (in panel (c)). Supplementary Figures 1 and 2 show more information on the variance and distribution of each connection for the HCP-YA data.  In total, we successfully extract brain structural connectomes for $1065$ subjects from the latest release of the HCP-YA dataset.}

\vspace{-0.5cm}
\subsection{Genetic data preprocessing}
A total of 1,141 subjects in the HCP-YA were genotyped using Illumina Multi‐Ethnic Global Array (MEGA) Chip and three specialized Illumina chips relevant to neuroimaging studies: Psych chip, NeuroX chip, and Immunochip. We download the MEGA chip genotype data from the HCP-YA dbGAP repository and perform the following quality control procedures: excluding subjects with more than $10\%$ missing SNPs, removing SNPs with larger than $5\%$ missing values, minor allele frequency
smaller than $5\%$, and Hardy-Weinberg equilibrium p-value less than $1\times 10^{-6}$. The resulting genotype data is further processed by a linkage disequilibrium (LD)-based pruning with window size 100, step size 10 and $r^2$ threshold 0.2, which leaves $130,452$ SNPs in our analyses. To determine block structure among genotypes, following \cite{zhao2019structured}, we start from an initial SNP with a putative block of size 100 and consider sub-blocks by gradually decreasing block size from 100 until more than 50\% of elements in the corresponding correlation matrix have $r^2$ surpassing 0.02. This results in 130,452 SNPs being partitioned into $23,418$ SNP-sets. Finally, a total of $1010$ subjects who have both the connectome and genetic features are included in our study.

\section{Method}\label{sec:method}
\vspace{-0.3cm}
\subsection{Model formulation}\label{sec:model}
We now introduce the notations of our model under a general setup. For subject $i$ $(i=1,\dots,N)$, a brain connectome is  summarized under a graph $\cG_i=(\mathcal{V}_i,\mathcal{E}_i)$ with $\mathcal{V}_i$ being the nodes (i.e., a set of brain ROIs) and $\mathcal{E}_i$ being the edges connecting the nodes. We assume $\mathcal{V}_i$ is the same for all subjects and it covers $V$ ROIs across the whole brain. Let $\bfx_i$ denote  high dimensional features with a size of $P$. In our application, we consider imaging genetics association using whole genome SNPs, adjusting for age, gender and top ten genetic principal components. We will initially  model with $\bfx_i$ including only genetic variants, and will  discuss additional covariate adjustment at the end of the section.  To quantify brain connectome variation, we represent each $\cG_i$ by its uniquely determined $V\times V$ adjacency matrix $A_i$, where its $(v,v')$th element $a_{ivv'}$ measures the connectivity between brain regions $v$ and $v'$, and $a_{ivv'}=a_{iv'v}$, $1\leq v, v' \leq V$. Depending on how brain connectivity is summarized, the values of $A_i$ can fall into different data types, e.g., a binary value, an integer, or a real number. Similar to \cite{zhang2018network}, we assume the element-wise conditional expectation of matrix $A_i$ as $\Psi_i=\mathbb{E}(A_i\mid \bfx_i)$. We then consider the following generalized network response high dimensional regression to assess the genetic association on brain connectivity network
\vspace{-0.2cm}
\begin{eqnarray}
\label{eqn:model}
g(\Psi_i)=B_0+\mathcal{B}\times_3 \bfx^T_i.
\end{eqnarray}
Here $g(\cdot)$ is the canonical link function for each matrix entry. When the graph edge represents connection status (e.g., the existence of white matter tracts), we will set $g(\Psi_i)=\mbox{logit}(\Psi_i)$ to model a binary response matrix; and when it summarizes fiber connection strength, we will use $g(\Psi_i)=\Psi_i$.  The $V\times V$ matrix $B_0=(b_{0vv'})$ is the intercept, which is symmetric in our application with the $(v,v')$th element $b_{0vv'}$ indicating the mean connection between regions $v$ and $v'$. The $V\times V\times P$ tensor $\mathcal{B}=(\beta_{vv'p})$ is the coefficient which characterizes the genetic effects on  brain connectome variation. Tensor $\mathcal{B}$ is semi-symmetric under each frontal slice, which leaves a symmetric matrix $\mathcal{B}_{:,:,p}$ to capture the impact of genetic marker $p$ on network phenotype. Of note, model \eqref{eqn:model} is consistent with existing models \citep{wang2019common,zhang2018network} in the sense that it decomposes a network response into a shared component and an individual component. However, unlike previous works, we associate the network outcome with a larger scale of structured predictors. By uncovering genetic bases along the genome  and  brain sub-networks underpinned by each biomarker, {we hope to enhance our understanding of the genetic control on higher level brain structural configuration and  prioritize tractable sets of genetic variants and network endophenotypes. }

It is well accepted that brain network data often admit low rank representations \citep{hoff2002latent}. From the implementation perspective, in the presence of high-dimensional predictors, directly estimating the coefficient tensor in model \eqref{eqn:model} is intractable.  Inspired by existing literature on signal sub-graph  \citep{wang2019learning,relion2019network}, we employ a low rank representation on the first two dimensions of the coefficient tensor 
\vspace{-0.2cm}
\begin{eqnarray}
\label{eqn:model2}
\mathcal{B}_{:,:,p}=\bfu_p\otimes\bfu_p; \quad \bfu_p=(u_{1p},\dots,u_{Vp})^T; \qquad p=1,\dots,P,
\end{eqnarray}
where $\otimes$ is the outer product. 
The representation above induces an explicit biological interpretation as each element $u_{vp}$ describes the effect of genetic variant $p$ on the brain network passing through region $v$, allowing a direct mapping from genetic variants to clique graphs \citep{alba1973graph} within a network. {The clique subgraph is formed by full interconnections among a subset of nodes, and has both modeling advantages and relevant biological interpretations. Compared with an arbitrary subgraph which is usually represented as a $V \times V$ adjacency matrix, a clique subgraph only needs $V$ unknown parameters, greatly simplifying the subgraph representation. Moreover, it is common that a small group of brain regions are interconnecting and collaborating with each other to realize a certain brain function. A clique graph can flexibly represent/approximate such a network module; biologically, we assume each SNP affects not only one brain region, but a set of brain regions and their interconnections. }  On the other hand, compared with the alternative semi-symmetric rank-$K$ tensor decomposition \citep{sun2017store}, i.e., $\mathcal{B}=\sum_{k\in[K]}w_k\bfbeta_{1k}\circ\bfbeta_{1k}\circ\bfbeta_{2k}$; $\bfbeta_{1k}\in \mathbb{R}^V, \bfbeta_{2k}\in \mathbb{R}^P$, model \eqref{eqn:model2} contains more degrees of freedom to distinguish the effect of each SNP on the network phenotype. Meanwhile, it also forms a clear parameter basis to facilitate  a biologically plausible shrinkage that will be discussed in the following section.
Of note, it is straightforward to extend \eqref{eqn:model2} to a rank-$K$ decomposition. However, we claim that with a primary goal to discover genetic effects and the corresponding network endophenotypes, setting $K=1$ marginally sacrifices the estimation accuracy while allowing feasibility and  interpretability. On a less crucial note, we do not assume any structure for $B_0$ to guarantee enough flexibility to estimate the shared network component. 

\subsection{Network response structural shrinkage}\label{sec:NRSS}

 Based on models \eqref{eqn:model} and \eqref{eqn:model2}, each vector $\bfu_p$ characterizes the effect of an individual genetic variant over the brain network response. Given the general sparsity among genetic signals as shown in previous empirical studies \citep{satizabal2019genetic}, {it is biologically meaningful to consider a sparse signal setting among genome-wide effects by assuming a large number of SNPs having negligible impact on the variation of brain networks. Additionally, different brain white matter microstructures have shown to be linked with distinct genetic architectures \citep{zhao2021common,arnatkeviciute2021genome}, suggesting that each piece of genetic signal tends to associate with a subgraph rather than the whole brain network. This finding supports us in associating each genetic factor with a sub-network signature. Meanwhile, 
since both the genome and  brain 
network contain complex biological architectures, a proper incorporation of such structural information will likely improve the statistical power and result in better interpretability. For instance, there is a grouping structure along the genome in the form of SNP-sets where $P$ SNPs can be grouped into $Q$ sets based on genes, pathways or LD. We denote this mapping matrix as $M=(m_{pq}) \in \real^{P\times Q}$ with $m_{pq}=1$ indicating SNP $p$ belongs to SNP-set  $q$. {Extensive literature on SNP-set analyses has showed its advantage over individual-SNP analysis by aggregating diluted genetic signals to gain statistical power \citep{wu2011rare,zhao2019structured}. Following \cite{wu2010powerful}, we construct SNP-set in light of LD to augment coverage of the genome in our numerical studies. In a word}, by incorporating the SNP-set information, we could improve the selection accuracy via a sparse group selection within genetic predictors.}

Combining the points above, we propose a novel network response structural shrinkage (NRSS) prior to perform a \textit{duo}-selection on both risk genotypes and their corresponding brain sub-networks:
\vspace{-0.2cm}
\begin{align}\label{eqn:prior1}
\begin{split}
u_{vp} &= f_{q}^{1/2} g_v h_{vp};  \quad  p=1,\dots,P; \quad v=1,\dots,V, \text{  and} \quad m_{pq}=1;\\
f_q &\sim \mathcal{EXP}(\lambda_q^f); \qquad g_v \sim \mathcal{EXP}(\lambda^g_v); \qquad  h_{vp} \sim \mathcal{L}(\lambda_{vp}^h),
\end{split}
\end{align}
where $\mathcal{EXP}(\cdot)$ and $\mathcal{L}(\cdot)$ stand for the Exponential and Laplace distributions, respectively. As we can see from model \eqref{eqn:prior1}, to allow for a structural shrinkage at the SNP-set level, brain network level, and network-SNP pair level, we split $u_{vp}$ into three components $f_q, g_v$ and $h_{vp}$.  The prior $f_{q}$ forms a group shrinkage for SNP-set $q$ and $h_{vp}$ forms a within-group shrinkage. By grouping SNPs into SNP-sets, we aggregate the risk effects that can otherwise be buried for their small to modest individual effects. The shrinkage induced by $g_v$ describes the overall genetic effect on the connections involving brain node $v$.

On the phenotype side, motivated by previous structural shrinkage ideas  \citep{zhao2015bayesian,chang2018}, we also consider imposing a smoothing effect over the nodes connected by a known population-based brain network to enhance biological interpretability. The population-based network can be either estimated under the current study or obtained externally from other brain imaging databases. We denote this prior network by $\widetilde{\Psi}$, with each element indicating the existence of an established fiber pathway connection in a population of subjects. By assuming the genetic impact on the  sub-networks induced by the connected nodes in $\widetilde{\Psi}$ admits a correlated effect, we adopt a recently developed structurally informative prior \citep{chang2018} for our sub-network shrinkage parameters $\bflambda^g=(\lambda^g_1,\dots,\lambda^g_V)^T$  as
\vspace{-0.2cm}
\begin{align}\label{eq:region1}
\bflambda^g &\sim \mathcal{LN}(\boldsymbol{\psi},\nu \Omega^{-1}),
\end{align}
where $\mathcal{LN}$ represents a Log-Normal (LN) distribution with vector $\boldsymbol{\psi}$ determining the overall sparsity of $\bflambda^g$, and $\nu$ controlling the mean concentration. Precision matrix $\Omega_{V\times V}$ captures the conditional dependency among $\bflambda^g$ with off-diagonal elements $\Omega(v,v')=-\omega_{vv'}=-\omega_{v'v}$ if $0\leq v\neq v'\leq V$, and $\Omega(v,v)=1+\sum_{v\neq v'}\omega_{vv'}$ to ensure positive definiteness. We extract its  upper diagonal elements as a vector $\bfomega$, and impose the following graph dependent prior 
\begin{align}\label{eq:region2}
\pi(\boldsymbol{\omega}) \propto |\Omega|^{-1/2} \prod_{v<v',\widetilde{\Psi}_{vv'}=1}\omega_{vv'}^{\tau-1} e^{-\eta \omega_{vv'}}\mbox{I}(\omega_{vv'}>0) \prod_{v<v',\widetilde{\Psi}_{vv'}=0} \delta_0(\omega_{vv'}),
\end{align}
where $\delta_0(\cdot)$ and $\mbox{I}(\cdot)$ are a Dirac delta function centered at 0 and  the indicator function; $\tau$, $\eta$ are shape and rate parameters from the Gamma distribution. Prior \eqref{eq:region2} induces a positive correlation between $\lambda_i^g$'s for the connected nodes in $\widetilde{\Psi}$ and leaves the rest independent, matching our objective to encourage correlated genetic effects among connected nodes in $\widetilde{\Psi}$. It is worth noting that when the existing brain topological structure is not applicable, we can easily replace priors \eqref{eq:region1} and \eqref{eq:region2} with a noninformative LN prior for $\bflambda^g$, and the two scenarios will be in high agreement when $\widetilde{\Psi}$ is super sparse. We suspect that  a large set of brain sub-networks only weekly associate with a genetic variant. If so, this prior is critical to aggregate phenotypic signals, and contributes to more interpretable results.   Finally, to adjust for non-genetic covariates such as age, gender, and the top genetic principal components in \eqref{eqn:model}, we can easily impose minimal shrinkage on each covariate by directly setting small $\bflambda^g$, large $\lambda_{vp}^h$ and $f_q=1$. We follow this strategy in our real data analyses.

\subsection{Operating properties}\label{sec:operating}
The proposed NRSS prior facilitates a biologically appealing shrinkage on the associations between high-dimensional SNPs and brain structural networks. Here, we investigate some additional operating characteristics of the proposed NRSS prior.

\begin{prop} \label{pro:dependency}
Assume $v_1$, $v_2$, $v_3$, and $v_4$ are four different nodes (i.e., brain regions). The prior correlations among the magnitude of entries in the coefficient tensor $\mathcal{B}$ are given as
\begin{align*}
    Cor (\log |\beta_{v_1v_2p}|,\log |\beta_{v_1v_2p'}|) &= \begin{cases}
    1, & p = p',\\
    \frac{\sigma_f^2 + 2\sigma_g^2}{\sigma_f^2 + 2\sigma_g^2 + 2\sigma_h^2}, &  p \neq p'; \exists q: m_{pq}=m_{p'q}=1,\\
    \frac{2\sigma_g^2}{\sigma_f^2 + 2\sigma_g^2 + 2\sigma_h^2}, & p \neq p';	\nexists q: m_{pq}=m_{p'q}=1. \end{cases}\\
    Cor (\log |\beta_{v_1v_3p}|,\log |\beta_{v_2v_3p'}|) &=  \begin{cases}
    \frac{\sigma_f^2 + \sigma_g^2 + \sigma_h^2}{\sigma_f^2 + 2\sigma_g^2 + 2\sigma_h^2}, & p = p',\\
    \frac{\sigma_f^2+\sigma_g^2}{\sigma_f^2 + 2\sigma_g^2 + 2\sigma_h^2}, &  p \neq p'; \exists q: m_{pq}=m_{p'q}=1,\\
    \frac{\sigma_g^2}{\sigma_f^2 + 2\sigma_g^2 + 2\sigma_h^2}, & p \neq p';	\nexists q: m_{pq}=m_{p'q}=1. \end{cases}\\
    Cor (\log |\beta_{v_1v_2p}|,\log |\beta_{v_3v_4p'}|) &= \begin{cases}
    \frac{\sigma_f^2}{\sigma_f^2 + 2\sigma_g^2 + 2\sigma_h^2}, & \exists q: m_{pq}=m_{p'q}=1,\\ 0, & \nexists q: m_{pq}=m_{p'q}=1. \end{cases}
\end{align*}
\end{prop}

First, the unique construction of NRSS induces a conditional correlation structure on the magnitude of elements in the coefficient tensor.
Suppose the shrinkage parameters are fixed homogeneously, i.e., $\lambda_q^f \equiv \lambda^f$, $\lambda_v^g \equiv \lambda^g$, and $\lambda_{vp}^h \equiv \lambda^h$, and let $\sigma_f^2$, $\sigma_g^2$, and $\sigma_h^2$ be the (conditional) prior variances of $\log f_q$, $\log g_v$, and $\log |h_{vp}|$.
Then Proposition \ref{pro:dependency} displays the dependency among the log-scale magnitudes of association effect inside $\mathcal{B}$. As we can see, our proposed prior assigns a larger correlation  on effects for connections within the same clique graph (e.g. $\log(\beta_{|v_1v_3p|})$ and $\log(\beta_{|v_2v_3p|})$), and SNPs within the same SNP-set (e.g. $\exists q: m_{pq}=m_{p'q}=1$).
In addition, if two nodes  $v_1$ and $v_2$ are linked within $\widetilde{\Psi}$, then $\lambda_{v_1}^g$ and $\lambda_{v_2}^g$ are positively correlated, introducing  correlation between $g_{v_1}$ and $g_{v_2}$ and thus imposing correlations to the log-coefficients.

\begin{figure}[t]\centering
    \includegraphics[width=4in]{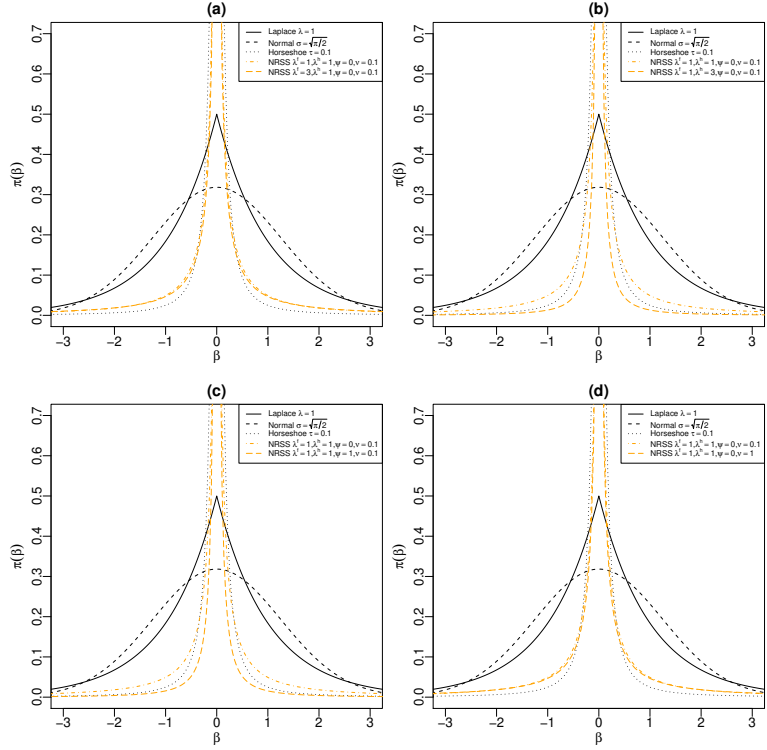}
  \caption{Comparison of the marginal density of individual coefficient in $\mathcal{B}$ under different prior shrinkage parameters, and with other shrinkage priors including Laplace, Normal and Horseshoe. }\label{fig:op}
\end{figure}

To further assess the operating properties, we compare the marginal density of each individual coefficient in $\mathcal{B}$ with other shrinkage priors including Laplace, Normal and the popular Horseshoe \citep{carvalho2010horseshoe} in Figure \ref{fig:op}. Figures \ref{fig:op}(a) and \ref{fig:op}(b) show how the marginal density varies with different values of $\lambda^f$ and $\lambda^h$.
Our marginal prior density of coefficient resembles the shape of the horseshoe prior, which has a higher peak at the origin and heavier tails compared to the Normal and Laplace priors.
Figure \ref{fig:op}(c) plots the marginal density of the regression coefficients for different values of $\psi$ with $\nu$ fixed.
Clearly, larger values of $\psi$ encourage greater shrinkage on the imaging genetic associations.
In addition, if $\nu = 0$, we have $\lambda_v^g = \psi$
resulting in a fixed Laplace shrinkage for $\bfg$. When $\nu > 0$ we could obtain a more non-concave log-density as shown in Figure \ref{fig:op}(d), which leads to a desired adaptive property with which smaller coefficients receiving relatively more shrinkage and larger ones receiving less shrinkage. 

\subsection{Estimation}\label{sec:EM}
Given the high-dimensional nature of our model, developing an efficient estimation algorithm  is critical to ensure its practical use. Since the marginal likelihood is close to intractable to optimize,  we employ an efficient EM algorithm to perform the posterior computation. We take an identity link for model \eqref{eqn:model} under Gaussian errors for now, and assign the error variance $\sigma^2$ a noninformative inverse gamma prior, $\sigma^2 \sim \mathcal{IG}(1/2,1/2)$.  
We are interested in estimating $\bfPhi=(\bff,\bfg,H,B_0, \bfalpha,\sigma^2)$, with $\bff = (f_1,...,f_Q)^T$, $\bfg =(g_1,...,g_V)^T$, $H=(h_{vp}) \in \real^{V \times P}$, and $\bfalpha = (\alpha_1,...,\alpha_V)$ with  $\alpha_v = \log \lambda_v^g$.

To find the maximum a posteriori (MAP) estimator for $\bfPhi$, we consider $\Omega$ to be a latent factor and marginalize it out. The loglikelihood becomes
\vspace{-0.2cm}
\begin{align*}
    \log \pi(\cA|B_0,\cB,\sigma^2) = C -\frac{NV(V-1)}{4} \log \sigma^2 - \frac{1}{2\sigma^2} \sum_i \sum_{v'>v} \left(a_{ivv'} - b_{0vv'} - \sum_{p=1}^P u_{vp} u_{v'p} x_{ip} \right)^2,
\end{align*}
where $\cA=\{A_1,\dots,A_N\}$. We also have the  logarithmic prior density for $(\bff, \bfg, H,\sigma^2)$
\vspace{-0.2cm}
\begin{align*}
    \log \pi(\bff,\bfg,H,\sigma^2|\bflambda^g) = C - \sum_q \lambda^f f_q + \sum_v (\log \lambda_v^g - \lambda_v^g g_v) - \sum_{v,p} \lambda^h |h_{vp}| - \frac{3}{2} \log \sigma^2 - \frac{1}{2\sigma^2},
\end{align*}
and that for the hyper-parameters $(\bfalpha, \bfomega)$ 
\begin{align*}
    \log \pi(\bfalpha,\bfomega) = C - \frac{(\bfalpha - \bfpsi)^T \Omega (\bfalpha - \bfpsi)}{2\nu}  + \sum_{\substack{v<v', \widetilde{\Psi}_{vv'}=1}} ((\tau-1) \log \omega_{vv'} - \eta \omega_{vv'}) - \sum_{\substack{v<v', \widetilde{\Psi}_{vv'}=0}} \delta_0(\omega_{vv'}).
\end{align*}
Suppose at iteration $t$,  we have $\bfPhi^{(t)}$.
Then at $t+1$, the EM algorithm updates parameters in $\bfPhi$ by maximizing the following objective function: $Q_t(\bfPhi) = \mbE \log  \pi(\bfPhi|\bfPhi^{(t)},\cA)$,
where the expectation is with respect to the conditional posterior distribution of $\bfomega$ with the other model parameters set to the values from the previous iteration.
Since the conditional distribution of $\bfomega$ only depends on $\bfalpha$, we have
\begin{align*}
    Q_t(\bfPhi) = C + \log \pi(\cA|B_0,\cB,\sigma^2) + \log \pi(\bff,\bfg,H|\bflambda^g,\sigma^2) + \mbE ( \log \pi(\bfalpha,\bfomega)|\bfalpha^{(t)}),
\end{align*}
with a constant $C$. Based on the above $Q_t$, we iteratively apply the following  E-step and M-step until convergence.  We also summarize this procedure in Algorithm \ref{alg:NRSS}.

\paragraph{E-step.}
The conditional posterior distributions of $\omega_{vv'}$'s are independently Gamma distributed with parameters $\tau$ and $\eta+\frac{\left(\alpha_v-\alpha_{v'} \right)^2}{2\nu}$ if $\widetilde{\Psi}_{vv'}=1$. Therefore, we have
\begin{align}
\omega_{vv'} \leftarrow \frac{2\nu \tau \widetilde{\Psi}_{vv'}}{ 2\nu \eta + ( \alpha_v -\alpha_{v'} )^2}, \qquad v\neq v'. \label{eq:Estep}
\end{align}
Procedure \eqref{eq:Estep} reflects how the partial correlations rely on the shrinkage parameters, and it also shows the strength of correlations imposed by the NRSS prior can be adaptive to the data.
This property can be particularly important when the observed data have connection patterns that are not fully consistent with the pre-specified population connectivity $\widetilde{\Psi}$.

\paragraph{M-step.}
In this step, we optimize the EM objective function by alternatively updating the parameters $\bff$, $\bfg$, $H$, $B_0$, $\bfalpha$, and $\sigma^2$.
To update $\bff$, $\bfg$, and $H$, we use the Iterative Shrinkage-Thresholding Algorithm in \cite{Beck2009}, which is a proximal gradient descent algorithm. 
We update $\bfalpha$ by the quasi-Newton search, and $B_0$ and $\sigma^2$ by closed form update formulas. Note that  $\bfu_v = g_v \bff_0 \circ \bfh_v$ where $\bff_0 = \diag(f_{q(1)}^{1/2}, \dots, f_{q(P)}^{1/2})$. To update $g_v$ and $\bfh_v$, we minimize
\begin{align*}
    \frac{g_v^2}{2} \bfh_v^T R_v \bfh_v - g_v \bfh_v^T \bfs_v + \lambda_v^g g_v + \lambda^h \sum_p |h_{vp}|, \qquad \text{subject to } g_v \ge 0;\\
    R_v = \bff_0 \bff_0^T \circ \sum_{v' \neq v} \bfu_{v'} \bfu_{v'}^T \circ \sum_i \bfx_i \bfx_i^T / \sigma^2,
    \text{ and }\bfs_v = \bff_0 \circ \sum_{v' \neq v} \bfu_{v'} \circ \sum_i \bfx_i (a_{ivv'} - b_{0vv'}) / \sigma^2.    
\end{align*}
This becomes a lasso problem with respect to $\bfh_v$, and we update $\bfh_v$ by a proximal gradient descent algorithm:
\begin{align*}
    \bfh_v \leftarrow \prox_{\lambda^h,l_v}(\bfh_v - l_v (g_v^2R_v \bfh_v - g_v \bfs_v)), \quad v=1,\dots,V,
\end{align*}
where $\prox_{\lambda,t}(\bfx) = \argmin_\bfz \left( \frac{1}{2t} \|\bfz-\bfx\|_2^2 + \lambda \|\bfz\|_1 \right)$ is the proximal operator associated with the lasso penalty.
Note that this operator has an analytic solution; if we let $\widetilde{\bfx} = \prox_{\lambda,t}(\bfx)$, we have $\widetilde{x}_j = (1-\lambda t/|x_j|)_+ x_j$ for all $j$.
Here, $l_v$ is the step size and can be determined by a backtracking line search method.
Since $R_v$ a is a $P \times P$ matrix, this step requires $O(P^2V)$ operations.
Similarly, we update $g_v$ with
$g_v \leftarrow (\bfh_v^T \bfs_v-\lambda_v)_+ / \bfh_v^T R_v \bfh_v$.

In order to update $\bff = (f_1,\dots,f_Q)^T$, note that
\begin{align*}
    \sum_{p=1}^P u_{vp} u_{v'p} x_{ip} = g_v g_s \sum_{q=1}^Q f_q \sum_{p:q(p)=q} h_{vp} h_{v'p} x_{ip}.
\end{align*}
Therefore, $f_q$ can also be updated by minimizing $ \frac{1}{2} \bff^T R_f \bff - \bfs_f^T \bff + \lambda^f \sum_q f_q$,
subject to $f_q \ge 0$ where $R_f= \sum_{i,v,v'} \bfy_{ivv'}\bfy_{ivv'}^T / \sigma^2$, $\bfs_f= \sum_{i,v,v'} (a_{ivv'} - b_{0vv'}) \bfy_{ivv'} / \sigma^2$ 
with $\bfy_{ivv'} = (y_{ivv'1},\dots,y_{ivv'Q})^T,  y_{ivv'q} =g_v g_{v'} \sum_{p:q(p)=q} h_{vp} h_{v'p} x_{ip}.$
To find the solution, the proximal gradient descent algorithm is used again: $
    \bff \leftarrow \prox_{\lambda^f,l'}^+(\bff - l'(R_f\bff - \bfs_f)),
$
where $\prox_{\lambda,t}^+(\bfx) = \argmin_{\bfz \ge \bfzero} \left( \frac{1}{2t} \|\bfz-\bfx\|_2^2 + \lambda \|\bfz\|_1 \right)$ with step size $l'$ determined by the backtracking line search. 

To update $\bfalpha$, the Newton search direction is given by $\bfd = -R_a^{-1} \bfs_a,$ where $R_a = \Omega/\nu + \mathcal{D}_{\bfg} \mathcal{D}_{e^{\bfalpha}}$ and $\bfs_a = \Omega \bfalpha /\nu - (1+\psi/\nu) \bfone + \mathcal{D}_{\bfg} e^{\bfalpha}$. Here, $e^{\bfalpha} = (e^{\alpha_1},\dots,e^{\alpha_V})^T$ and $\mathcal{D}_{\bfa} = \diag(\bfa)$ for any vector $\bfa$. 
Finally, the closed form updates for the intercept $B_0$ and random error variance $\sigma^2$ follow
$b_{0vv'} \leftarrow  \bar{a}_{vv'} - \bar{\bfx}^T (\bfu_v \circ \bfu_{v'})$,
where $\bar{a}_{vv'} = \frac{1}{N} \sum_i a_{ivv'}$ and $\bar{\bfx} = X^T \bfone/N$, and
$\sigma^2 \leftarrow \frac{2}{NV(V-1)} \sum_{i,v<v'} \left(a_{ivv'} - b_{0vv'} - \sum_p u_{vp} u_{v'p} x_{ip} \right)^2$.

The above algorithm can be easily modified to accommodate other link functions, say a logit link.
The proximal gradient descent method only requires an evaluation of likelihood gradient. Therefore, we can model different types of network connections as long as the likelihood is differentiable and smooth \citep{Beck2009}.

\begin{algorithm}[ht]
\caption{EM Algorithm for NRSS}
   \label{alg:NRSS}
    Initialize $\bff$, $\bfg$, $H$, $B_0$, $\bfalpha$, $\Omega$, and $\sigma^2$;
    
    \Repeat{convergence}{
        \For{$v\gets1$ \KwTo $V-1$}{
            \For{$v'\gets v+1$ \KwTo $V$}{
                $\omega_{vv'} \leftarrow \frac{2\nu \tau \widetilde{\Psi}_{vv'}}{ 2\nu \eta + ( \alpha_v -\alpha_{v'} )^2}$;
                
                $b_{0vv'} \leftarrow \bar{a}_{vv'} - \bar{\bfx}^T (\bfu_v \circ \bfu_{v'})$;
            }
        }
        
        \For{$v\gets1$ \KwTo $V$}{
            $\bfh_v \leftarrow \prox_{\lambda^h,l_v}(\bfh_v - l_v (g_v^2R_v \bfh_v - g_v \bfs_v))$;
        
            $g_v \leftarrow (\bfh_v^T \bfs_v-\lambda_v)_+ / \bfh_v^T R_v \bfh_v$;
        }
        
        $\bff \leftarrow \prox_{\lambda^f,l'}^+(\bff - l'(R_f\bff - \bfs_f))$;
    
        $\sigma^2 \leftarrow \frac{2}{NV(V-1)} \sum_{i,v<v'} \left(a_{ivv'} - b_{0vv'} - \sum_p u_{vp} u_{v'p} x_{ip} \right)^2$;
        
        Update $\bfalpha$ by one step of the Newton method;
    }

\end{algorithm}


\section{Real Data Analysis of HCP-YA Data}
\label{sec:realdataHCP-YA}

We apply our network response model to HCP-YA data based on the extracted imaging genetic features from 1010 subjects described in Section \ref{sec:data}. Along with the $130,452$ SNPs, we also adjust for age, gender, and the first 10 genetic PC scores. {To assess the out-of-sample performance, we first randomly split the whole dataset into training, test and validation sets with sample sizes $506$, $252$, and $252$, respectively. Then we conduct a greedy search over different hyper-parameter combinations on $\lambda^h$, $\lambda^f$, $\boldsymbol{\psi}$ and $\nu$ to get a rough range for each parameter in NRSS by fitting the model to the training set.  Based on the predicted performance in the validation set, we narrow down to nine hyper-parameter combinations. Note that we use all HCP-YA subjects to calculate $\widetilde{\Psi}$: for a region pair $(i,j)$, if all subjects have a white matter connection between the two regions, we set $\widetilde{\Psi}$(i,j) = 1. Supplementary Figure 1(b) shows the final $\widetilde{\Psi}$ used. Next, we run our model using each of the selected combinations of hyper-parameters with 40 random splits of the whole HCP-YA dataset (same splitting as before, $506$ for training, $252$ for validation, and $252$ for testing).  The computing is done using Yale's High Performance Computing. For each subject, with one CPU core (comparable with Intel Xeon 2.5 GHz) and 20 GB RAM, it takes about 24 hours to process the subject's brain imaging data to extract the brain network. The NRSS model fitting usually can be done in 24 hours with one CPU core and 80 GB RAM.

To stabilize the final selection of SNP biomarkers and their corresponding
sub-networks, results from different random splits and parameter combinations
are synthesized using the stability selection method \citep{meinshausen2010stability}. Note that stability selection is  essentially  a  boosting  method  and  requires  random  subsampling. The total 40*9 random splits creates 360 different training datasets for this boosting procedure. To summarize the results,   we calculate the selection probability of each element in $\mathcal{B}$ as $\Pi^{\mathcal{B}}$, and threshold $\Pi^{\mathcal{B}}$ using a threshold $\pi_{thr}$ (denoted as $\mathcal{I}^{(\mathcal{B}>\pi_{thr})}$). If there is at least one non-zero in  $\mathcal{I}^{(\mathcal{B}>\pi_{thr})}_{:,:,p}$ for a certain $p$, then SNP $p$ is considered to be a significant biomarker. The $\pi_{thr}$ is selected according to the upper bound of falsely selected variables in equation (9) in \citep{meinshausen2010stability} so that we have an upper bound of FDR (false discovery rate) around $0.3$ for SNPs. Although the test and validation sets are not directly involved in identifying signal SNPs, they are used to evaluate the predictive performance of NRSS and the convergence of the model fitting. Overall, in the test and validation datasets, our model can achieve a high correlation $r = 0.90$ for predicting the network response. }

\begin{figure}[H]\centering
    \includegraphics[width=5.5in]{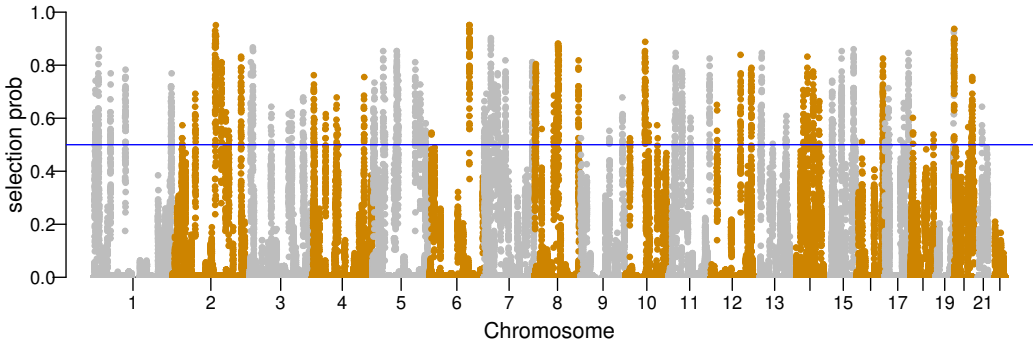}
  \caption{\small{Manhattan plot for the selection probability of all the genetic variants in the HCP-YA application.}}\label{fig:snp}
\end{figure}  

\begin{figure}[h]
\centering
\setlength\tabcolsep{2 pt}
\begin{tabular}{cc}
\includegraphics[width=6cm]{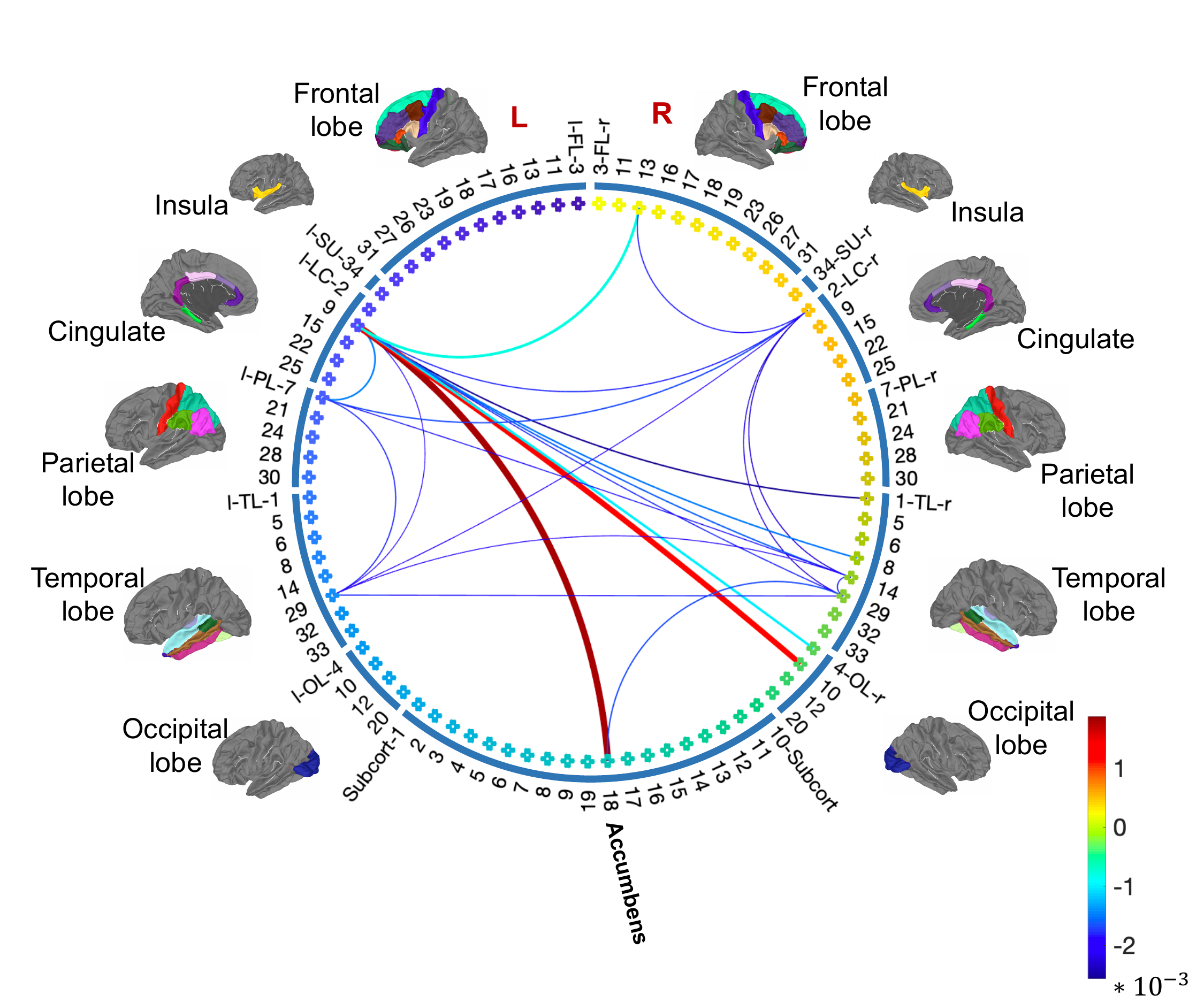}&
\includegraphics[width=6cm]{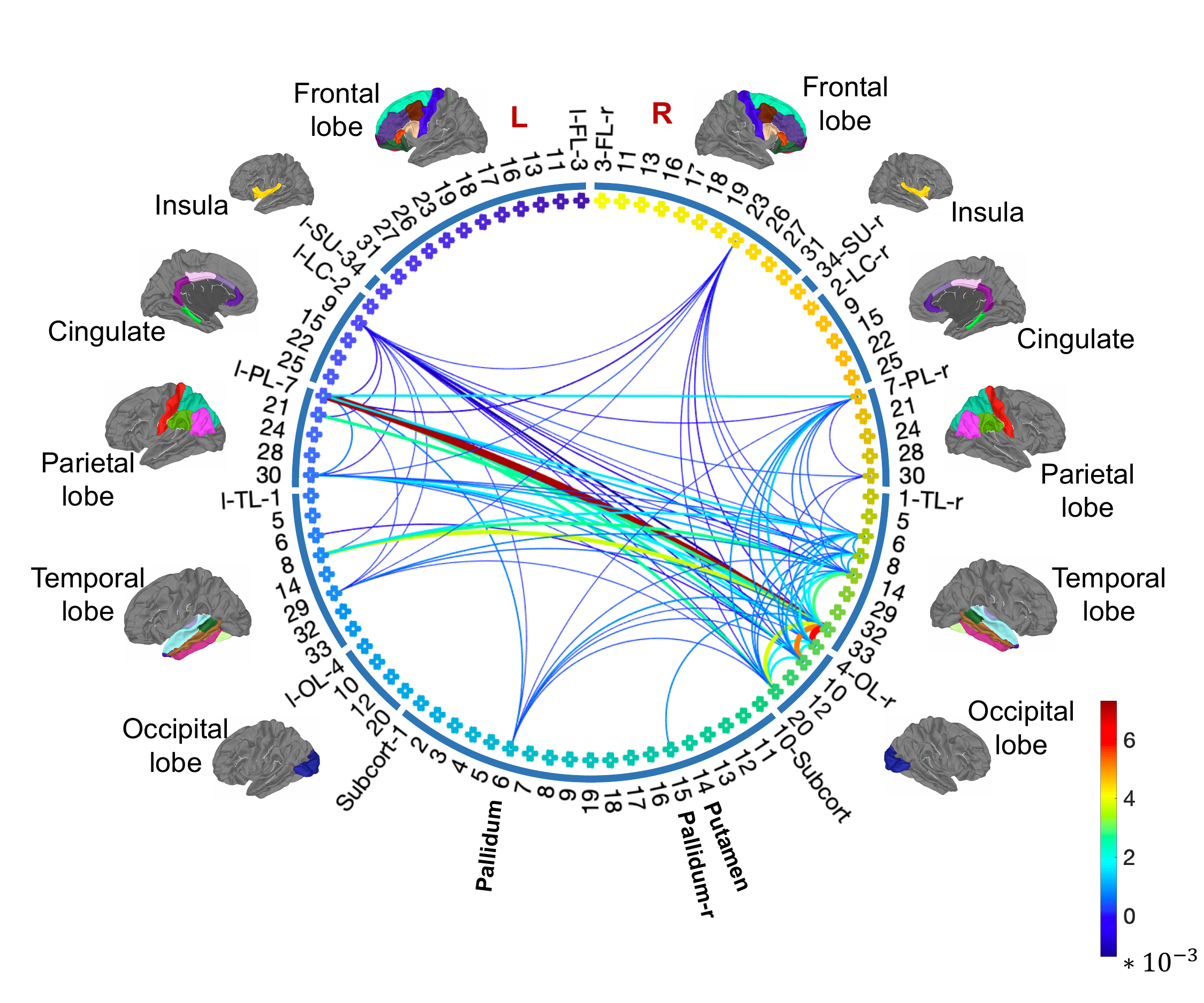}\\
     rs75712269 & rs630683 \\
    \multicolumn{2}{c}{
    \includegraphics[width=6cm]{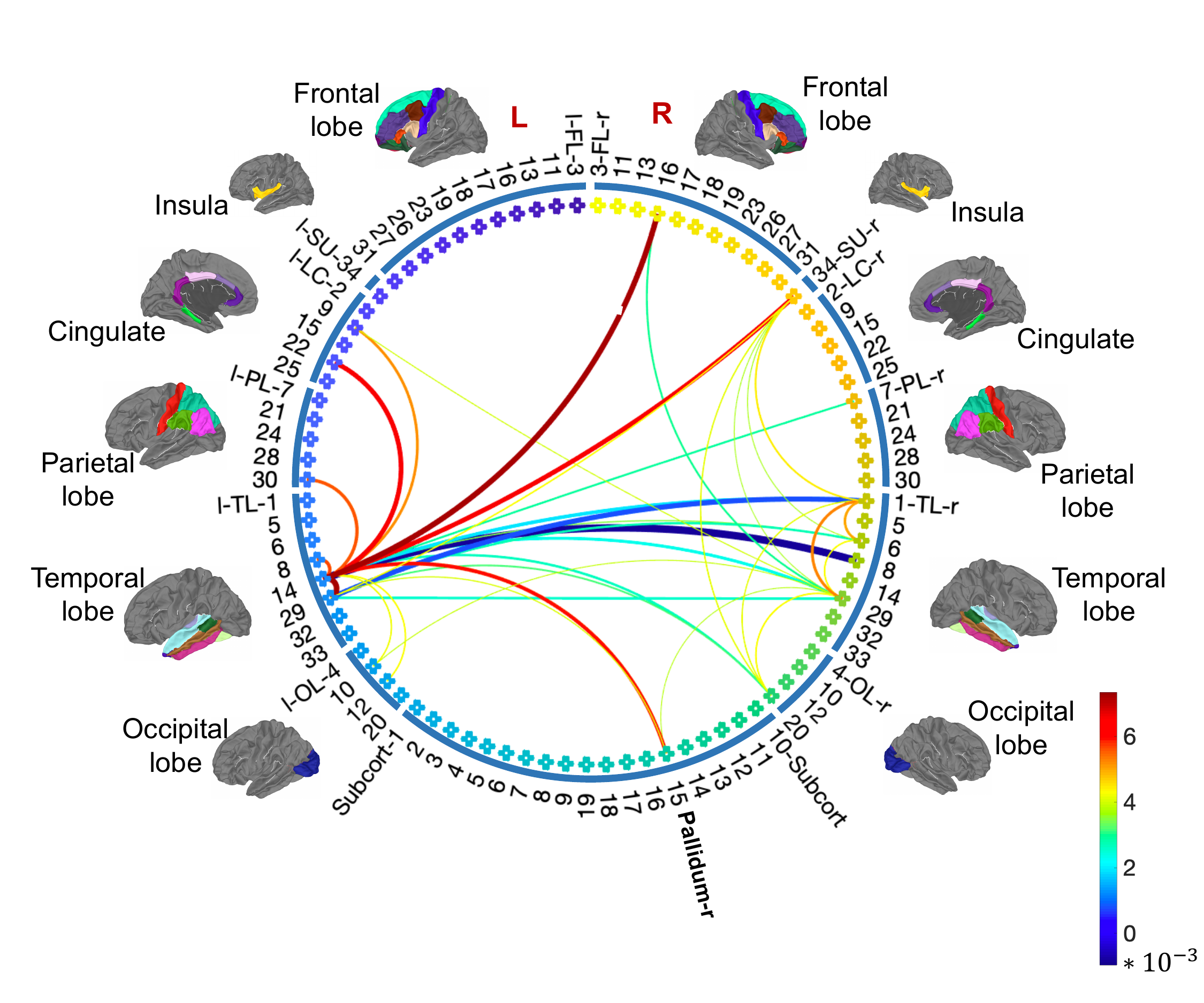}}\\
    \multicolumn{2}{c}{rs79738048}
\end{tabular}
    
    \caption{Top three genetic signals rs75712269, rs630683 and rs79738048, their impacted brain sub-networks (i.e., the corresponding imaging genetic associations ${\mathcal{B}}_{:,:,p}$). }
    \label{fig:coeffexp}
\end{figure}

We show the selection probability for each SNP in a Manhattan plot in Figure \ref{fig:snp}, representing the chance of SNP $p$ being important across our experiment runs. Based on the figure, we notice the peaks, which indicate the informative genetic variants,  spread out the whole genome. Based on a $0.5$ threshold, we identify $1860$ genetic signals showing significant associations with 3549 unique structural connections. The top twelve SNPs include rs1257174, rs192686405, rs75712269, rs7767478, rs9491189, rs630683, rs79738048, rs1591705, rs781738, rs17053023, rs159679 and rs204681; { among all the identified gene variants, 891 of them are the independent leading ones under an LD $r^2<0.1$.} {Figure \ref{fig:coeffexp} shows three examples of the estimated  ${\mathcal{B}}_{:,:,p}$ for SNPs rs75712269, rs630683 and rs79738048, where the color and line width of each connection indicate the strength of SNP's effect on the brain network.}  We further map all the selected SNPs to the closest genes they belong to and obtain 427 unique genes. After performing a functional annotation to the sources, including Kyoto Encyclopedia of Genes and Genomes (KEGG), Reactome and Gene Ontology (GO), we obtain one KEGG pathway, 13 GO terms and 2 Reactome pathways after multiple comparison adjustment. {A complete list of selected SNPs with their selection probabilities, independent leading  variants, annotated genes and pathways are provided in the supplementary materials. }

Taking a closer investigation on the enriched GO terms and pathways, we conclude our results offer a strong biological interpretability. For instance, the identified KEGG pathway synaptic vesicle cycle represents a series of steps that tightly regulate exo- and endocytosis of vesicles in cortical and hippocampal neurons of the brain \citep{sudhof1995synaptic}. The pathway itself forms the basis for neurotransmitter release and plays a primary role in driving cognitive processes such as learning and memory \citep{kennedy2016synaptic,powell2006gene}.
Among the identified GO terms and Reactome pathways, five of the GO terms (GO:0015171, GO:0005416, GO:0005342, GO:0046943, GO:0006865) and one Reactome pathway (R-HSA-428559) are all related to amino and other common acid transport activities. These acid transmitters are known to provide the majority of excitatory and inhibitory neurotransmission in the nervous system, and transport across blood–brain barriers which are crucial to maintain the microenvironment of white matter tracts \citep{choi2016effects,villabona2019evolving,butt2014neurotransmitter}. Another Reactome pathway (R-HSA-9033241) is related to peroxisomal proteins, and such proteins have been shown to highly impact the normality of white matter in the nervous system \citep{kassmann2014myelin,hulshagen2008absence}.

\begin{figure}[H]
\centering
\begin{tabular}{c}
\includegraphics[width=12cm]{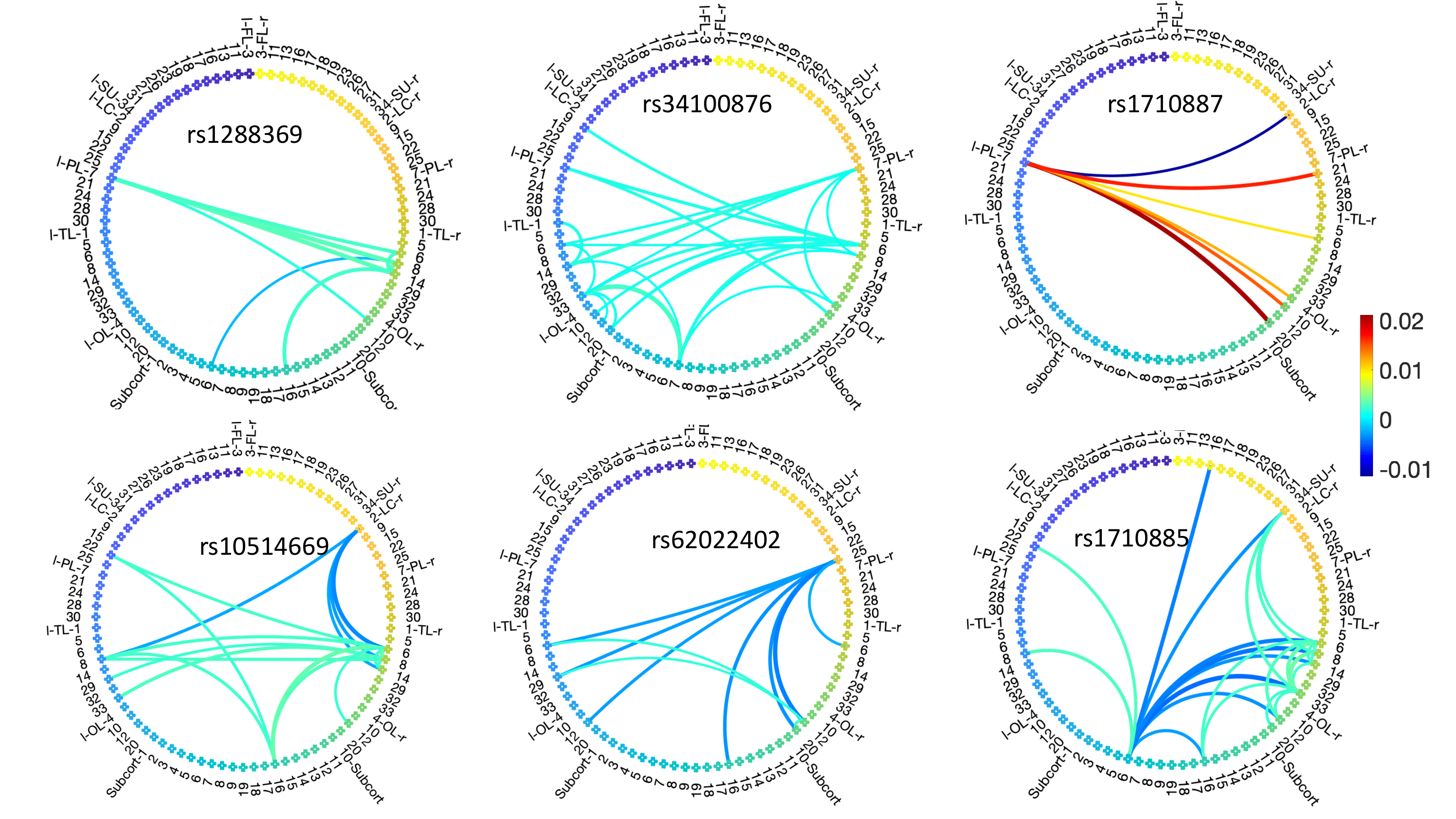}
\end{tabular}
    
    \caption{Genetic-network association ${\mathcal{B}}_{:,:,p}$ corresponding to six of the selected SNPs in the discovered synaptic vesicle cycle pathway. The full set of results can be found in the supplementary materials.  }
    \label{fig:keggexample}
\end{figure}

Taking the synaptic vesicle cycle pathway as an example, we display the genetic-network associations of all the selected SNPs within that pathway in the supplementary materials and present a subset in Figure \ref{fig:keggexample}. From these results, we see the most frequently selected subcortical ROIs include subcort- 6, 7 and 16, corresponding to left pallidum, left hippocampus, and right hippocampus, respectively. If we pay closer attention to all the brain connections corresponding to the selected SNPs in this pathway, we can conclude that the cross-hemispheric connections and cortical to sub-cortical connections dominate over other types of connections with $41.5\%$ cross-hemispheric connections, $22.3\%$ cortical to subcortical connections, and only $36.2\%$ other types of connections (cortical to cortical and sub-cortical to sub-cortical). 
From neuroscience literature, we know that the hippocampus plays critical roles in memory, navigation, and cognition \citep{moser1998functional, knierim2015hippocampus,lisman2017viewpoints}, and cross-hemispheric structural connections are found to be associated with high-level cognition abilities, e.g., language-related abilities \citep{zhang2019tensor}, and creativity \citep{durante2018bayesian}.  

\begin{table}[ht]
	\centering
		\caption{\small{Significant cross-tissue cis-eQTL results obtained from UKBEC brain database}}\label{table:eqtl}
	\newsavebox{\habox}
	\begin{lrbox}{\habox}
		\begin{tabular}{ccccccccc}
			\toprule
SNP& Chr &$P$-value&Regulated genes &&SNP& Chr &$P$-value&Regulated genes\\
		\cline{1-4}
		\cline{6-9}
rs301506	&	5	&	5.30E-09	&	AMACR	&	&	rs56078509	&	8	&	6.60E-06	&	HNF4G	\\
rs62365867	&	5	&	1.59E-08	&	LOC100131067	&	&	rs1320594	&	1	&	6.79E-06	&	SLC1A7	\\
rs1022855	&	14	&	6.60E-08	&	LRFN5	&	&	rs245343	&	5	&	1.01E-05	&	DHFR	\\
rs7146542	&	14	&	8.31E-08	&	LRFN5	&	&	rs17001357	&	4	&	1.27E-05	&	NAAA	\\
rs2287363	&	9	&	2.24E-07	&	PHYHD1	&	&	rs7778745	&	7	&	1.30E-05	&	DBNL	\\
rs4801595	&	19	&	5.45E-07	&	ZNF132	&	&	rs10768994	&	11	&	1.33E-05	&	SEC14L1,ALKBH3,LOC729799	\\
rs17170220	&	7	&	6.03E-07	&	BBS9	&	&	rs12547746	&	8	&	1.48E-05	&	TMEM70	\\
rs73347189	&	7	&	6.11E-07	&	MIOS	&	&	rs284574	&	2	&	1.50E-05	&	RPL37A	\\
rs465047	&	20	&	7.40E-07	&	CTSZ	&	&	rs9791117	&	5	&	1.70E-05	&	MSH3,DHFR	\\
rs424504	&	20	&	1.10E-06	&	NPEPL1	&	&	rs4478425	&	6	&	2.30E-05	&	TRDN	\\
rs4704706	&	5	&	1.32E-06	&	LOC100131067	&	&	rs13284665	&	9	&	2.39E-05	&	ENDOG,C9orf114	\\
rs6725601	&	2	&	1.70E-06	&	STK39	&	&	rs59746931	&	1	&	2.49E-05	&	KAZ,C1orf126	\\
rs56078509	&	1	&	5.90E-06	&	ZCCHC11	&	&	rs3743637	&	16	&	2.63E-05	&	TAF1C,ADAD2	\\
rs6935961	&	6	&	6.17E-06	&	HDDC2	&	&	rs917454	&	7	&	2.80E-05	&	KBTBD2	\\
\bottomrule	
	\end{tabular}%
	\end{lrbox}

\scalebox{0.72}{\usebox{\habox}}
\end{table}

Finally, we perform a brain tissue-specific expression quantitative trait loci (eQTL) analysis using eQTL data from UK Brain Expression Consortium (UKBEC) \citep{ramasamy2014genetic}. UKBEC
generated genotype and exon-specific expression data for 10 brain tissues from 134 neuropathologically healthy individuals. {Through their web server BRAINEAC (http://www.braineac.org/), we evaluate our identified SNPs on their alteration of  gene expression  for each of the 10 brain tissues as well as across all the tissues. Following \cite{yao2020regional}, we focus on the cis-effect, and consider the gene expression profiles located within 100kb centered by each of the SNPs.  Eventually, among the 1860 identified SNP variants, {we discover 1700 unique ones. The BRAINEAC server directly provides a cross-tissue cis-effect p-value along with tissue-specific ones. To further ensure robustness, we also compute a cross-tissue p-value by merging the p-values obtained from individual tissue via the Cauchy combination test \citep{liu2020cauchy}. The detailed information including tissue-specific cis-effect p-values, and cross-tissue  cis-effect p-value provided by the server and computed by the Cauchy combination  test are summarized in the supplementary materials. We present here the more conservative p-values from the combined test. A total of 1412 cis-eQTLs are significantly associated with cross-tissue brain traits under a 0.05 cutoff, and we can still keep 28 of them after Bonferroni correction under a threshold of $0.05/1700=2.94e^{-5}$ as shown in Table \ref{table:eqtl}. Among the 28 eQTLs, there are 31 unique genes with expression regulated. We also summarize the result based on the cross-tissue p-values from  BRAINEAC server  (Supplementary Table 1), which shows even more significant cis-eQTLs.  Meanwhile, we also summarize the significant eQTLs associated with each single brain tissue, and the numbers of significant ones range from 263 to 324 under a 0.05 cutoff, and 2 to 7 under a $2.94e^{-5}$ one for a single brain tissue (Supplementary Table 2). The big difference in the SNP cis-effects upon gene expression across all brain tissues and under a specific brain tissue is consistent with the fact that brain structural connectome reflects a complex inter-brain architecture which is very different from a tissue/node specific phenotype. The identified across-tissue eQTLs also indicate the role of our identified SNPs on molecular regulation of brain white matter pathways through gene expression, and provide useful insights into molecular mechanisms involved in connectome-related neurogenetic processes.} }

\section{Simulations}\label{sec:simulation}
To further evaluate the performance of the proposed method and compare it with existing alternatives, we conduct extensive simulations in light of our imaging genetic application. {Our simulations contain two studies: 1) we consider a moderate number of genetic features to make the overall computation feasible for the competing approaches; 2) we consider a large number of genetic features features under the same scale as our real data analysis. }

We start with Simulation 1. In each simulated data set, we generate the genetic variants $\bfx_i$ from the Hapmap projects phase III data released by the International HapMap Consortium \citep{international2010integrating}. With a sample size of 300, we randomly combine two haplotypes from the CEPH population to form the genotypes for each subject. Among the genotypes, we first determine their block structure. Starting from an initial SNP $p$ with a putative block of 100 SNPs, we sequentially consider subsets of SNPs $\{p,p+1,\dots,p+d\}$ with $d$ starting from 99 and gradually decreasing until 50\% of the elements in the $r^2$ matrix surpass a certain threshold. Controlling this threshold to allow for at least 20 blocks containing more than 100 SNPs, we randomly choose 20 sets  and thus 2000 SNPs as predictors. To mimic the sparse genetic effects, we assume $5\%$ (i.e., 100) SNPs are associated with brain connectivity variations, and we consider the following two genetic signal patterns -- \textit{Pattern 1 (P1)}: We randomly select two SNP-sets and then $50$ risk SNPs from each of the two sets; \textit{Pattern 2 (P2)}: We randomly select ten SNP-sets and then $10$ risk SNPs from each of the ten sets. Here, \textit{P1} represents a more concentrated signal pattern, while \textit{P2} creates diluted signals to challenge the detective power of our model. For the brain connectivity, we set $V=20$ and $V=70$ to mimic low and high imaging resolutions. To construct coefficient tensor $\mathcal{B}$, we consider two cases: 1) following the model assumption with $\mathcal{B}_{:,:,p}=\bfu_p \otimes \bfu_p$ (\textit{clique sub-graph}), and 2) violating the assumption with a more complex $\mathcal{B}_{:,:,p} = \sum_{r=1}^3 \bfu_{pr} \otimes \bfu_{pr}$ (\textit{non clique sub-graph}).
Under both settings, for each risk SNP $p$, the non-zero elements within $\bfu_p$ are randomly selected and each non-zero element is independently sampled from a Normal distribution with mean 0 and variance $0.5$.  When $V=20$, we maintain around $75\%$ of the elements in each $\bfu_p$ as non-zero, and when $V=70$ we decrease this percentage to around $50\%$. 
We ultimately generate the connectivity matrix $A_i$ under two different signal to noise ratios (SNRs) with $\sigma^2 = 0.1$ (SNR$\approx54$) and  $\sigma^2=0.5$ (SNR$\approx14$). 
{Overall, in Simulation 1 with a moderate number of genetic features, our simulations include 16 different scenarios, allowing a comprehensive evaluation of the proposed method.}  For each scenario, we generate 100 data sets and randomly split each data set into equally sized training, validation, and test sets. We fit the model with different  hyper-parameter combinations under the training set, select the optimal combination under the validation set, and  evaluate the  model performance under the test set. Another use of the training data is to summarize the prior connectivity matrix $\widetilde{\Psi}$ based on thresholding (i.e. $\widetilde{\Psi}(i,j) = 0$ if none of the subjects has a connection between nodes $i$ and $j$). {
Considering that in practice there may be cases when $\widetilde{\Psi}$ is not available or even incorrectly specified, we include a simulation with 30\% elements in $\widetilde{\Psi}$ mis-specified, and denote this new mis-specified population connection matrix as $\widetilde{\Psi}_v$. By implementing our method under both $\widetilde{\Psi}$ and $\widetilde{\Psi}_v$, we  assess the efficacy of such prior information and the robustness of our model to prior mis-specification. }

In terms of competing methods, given few models are directly applicable for a network response regression under relatively high dimensional predictors, we choose the following three univariate response predictive models that can accommodate high dimensional sparse signals: 
 a recent EM-based scalable Bayesian variable selection approach (EMSHS) \citep{chang2018}, Bayesian lasso (Blasso) \citep{park2008bayesian} and sparse group lasso (SGL) \citep{yuan2006model}. Here, EMSHS and SGL can achieve structural shrinkage/sparsity within SNPs. Since these algorithms are not designed to handle a network response, we transform the upper triangle elements of each matrix $A_i$ into a vector and perform element-wise regression. 
The R packages {\tt EMSHS}, {\tt monomvn} and  {\tt SGL} are used to fit these models, following the same training, validation and testing procedure. {Note that our model is implemented with correctly specified prior $\widetilde{\Psi}$ (NRSS) and mis-specified $\widetilde{\Psi}_v$ (NRSS$_v$).} Finally, to compare models, we adopt the following metrics: for the estimation and prediction accuracy, we compute the mean squared error (MSE) of coefficient tensor $\mathcal{B}$ (MSE$_{\mathcal{B}}$) and the mean
squared prediction error (MSPE) of brain connectivity $A_i$ (MSPE$_A$); for the variable selection accuracy, we provide the area under the curve (AUC) to evaluate imaging genetic signals captured by the zero and non-zero elements in $\mathcal{B}$ (AUC$_{\mathcal{B}}$) and to identify genetic biomarkers (AUC$_{snp}$) when at least one of the element in $\mathcal{B}_{:,:,p}$ is non-zero. We also summarize the computational cost for each method 
considered.

We summarize the Simulation 1 results in Table \ref{tab:modeleva}.  We can clearly see  our NRSS model outperforms the competing methods in almost all scenarios under each evaluation metric, showing NRSS's superiority in estimating imaging genetic associations, predicting connectome phenotypes, and identifying genetics and brain sub-network signals. More specifically, 
 among different settings, when the genetic signals are more concentrated (\textit{P1}),   higher accuracy in selecting both genetic biomarkers and their associated brain sub-networks are achieved under almost all the methods compared with the cases with more scattered signals (\textit{P2}). In addition, under the low resolution ($V=20$), all the methods achieve more accurate estimation for coefficient tensor (MSE$_{\mathcal{B}}$) and prediction for connectivity phenotype (MSPE$_{A}$) compared with the high resolution ($V=70$). At the same time, given the true numbers of signals in the two cases are comparable, we do not see obvious differences in AUC$_{\mathcal{B}}$ and AUC$_{snp}$ for feature selection. Moreover, high SNR helps all methods to identify risk genetic factors and their associated brain network endophenotypes. 

    In terms of model assumptions,  when our assumption on the structure of coefficient tensor is violated, we do see a slight deterioration of the NRSS's performance. However, lower accuracy in almost all the metrics is observed under a more complex sub-network structure since the relationship between genetics and brain connectomes becomes more challenging to uncover. Thus, though NRSS is constructed leveraging the clique graph structure, our model is feasible to quantify general effect patterns between SNPs and brain connectomes through this approximation. In addition, comparing the results of NRSS and NRSS$_v$ indicates that an incorporation of a properly specified prior connectivity matrix improves the model performance. Meanwhile, considering that NRSS$_v$ still substantially outperforms the remaining competing methods, we anticipate our method will be robust to this prior setting.

\begin{table}
	\centering
	\caption{\small{Results for Simulation 1 under different settings evaluated by mean squared error of $\mathcal{B}$ (MSE$_{\mathcal{B}}$), mean
squared prediction error of $A_i$ (MSPE$_A$), AUC for imaging genetic signal detection in $\mathcal{B}$ (AUC$_{\mathcal{B}}$), AUC for genetic biomarkers selection (AUC$_{snp}$), and computational cost. The Monte Carlo standard deviation is included in
the parentheses.}}	
	\label{tab:modeleva}
	\scalebox{0.60}{%
		\begin{tabular}{clccccccccccc}  \toprule 
			&  & \multicolumn{5}{c}{\textbf{\textit{Clique graph}}} && \multicolumn{5}{c}{\textbf{\textit{Non clique graph}}} \\ 
			\cmidrule{3-7} \cmidrule{9-13}
			Settings  & Model	
			& MSE$_{\mathcal{B}} \times 10^3$  & MSPE$_{A}$  &  AUC$_{\mathcal{B}}$  &AUC$_{snp}$  & Time(s)     
			&& MSE$_{\mathcal{B}}$ $\times 10^3$ & MSPE$_{A}$  &  AUC$_{\mathcal{B}}$   &AUC$_{snp}$ & Time(s)  \\ \cmidrule{1-7} \cmidrule{9-13}
            & 
			EMSHS           & 6.44(0.22)   & 6.53(0.59)    & 0.51 & 0.72  & 19.81  
			&& 20.99(0.51)  & 21.25(1.76)   & 0.51 & 0.72  & 20.83   \\	
			$P1,V = 20,$
			&Blasso          & 10.60(0.47)  & 14.37(1.06)   & 0.51 & 0.88  & 1603.05 
			&& 32.72(1.53)  & 45.71(4.31)   & 0.51 & 0.78 & 1717.76 \\
			$\sigma^2 = 0.1$
			&SGL             & 49.93(3.84)  & 84.33(12.95)  & 0.63 & 0.93 & 369.63  
			&& 150.42(9.44) & 253.45(36.28) & 0.64 & 0.77 & 645.86 \\
			& NRSS$_v$  &0.88(0.47)   &0.98(0.47)   & 0.92 &0.99  & 555.66  
			&& 15.27(0.75)  &13.95(1.1)   & 0.91  &0.95  &34.12 \\
					&{\bf NRSS} 		  & 0.11(0.07)   & 0.21(0.06)   & 0.98 & 0.99 & 92.68  
			&& 16.23(0.81)  & 17.15(1.71)   & 0.94 & 0.95 & 130.94 \\
			&            	&  &     &       &       &&  &     &       &       \\

			&
			EMSHS          & 6.52(0.21)   & 7.02(0.60)    & 0.51 & 0.71  & 18.26  
			&& 21.10(0.50)  & 21.76(1.79)   & 0.51 & 0.71  & 21.88  \\
			 $P1, V = 20,$
			&Blasso          & 10.76(0.55)  & 15.13(1.33)   & 0.51 & 0.86 & 1568.40  
			&& 33.02(1.48)  & 46.36(4.17)   & 0.51 & 0.77 & 2468.66 \\
			$\sigma^2 = 0.5$
			&SGL            & 49.87(3.51)  & 83.68(12.09)  & 0.61 & 0.88 & 496.40  
			&& 149.09(8.80) & 249.86(34.66) & 0.63 & 0.73 & 710.47  \\
			& NRSS$_v$  &1.20(0.41)   & 1.68(0.4) & 0.92 &0.99	  &540.79  
			&&  15.60(0.77) & 14.65(1.17)  & 0.91  & 0.95 &35.31  \\
						&{\bf NRSS}  	  & 0.89(0.52)   & 1.57(0.62)    & 0.96 & 0.99  & 140.40 
			&& 16.72(0.90)  & 18.17(1.84)   & 0.93 & 0.95  & 146.39 \\
			&            	&  &     &       &        &&  &     &       &       \\

			&
			EMSHS          & 9.61(0.25)   & 9.60(0.76)    & 0.51 & 0.79  & 121.05 
			&& 29.95(0.71)  & 29.67(2.21)   & 0.51 & 0.76  & 114.34  \\
			$P1, V = 70,$
			&Blasso          & 15.57(0.65)  & 21.23(1.39)   & 0.51 & 0.79 & 7165.94 
			&& 46.54(1.87)  & 63.65(4.12)   & 0.51 & 0.55 & 7173.83 \\
			$\sigma^2 = 0.1$
			&SGL            & 71.66(4.34)  & 122.40(14.77)  & 0.62 & 0.79  & 1953.32
			&& 206.67(10.78)& 354.38(39.03) & 0.63 & 0.56  & 3115.07 \\
			&NRSS$_v$  &0.20(0.11)  &0.31(0.13)   &0.99  & 1.00 & 1009.81  
			&&  22.90(0.94) & 20.82(1.57)  & 0.95  & 0.97 &506.19 \\
						&{\bf NRSS}        & 0.03(0.03)   & 0.14(0.05)    & 1.00 & 1.00   & 331.18  
			&& 23.36(0.95)  & 23.09(2.18)   & 0.97 & 0.97   & 473.46  \\
			&            	&  &     &       &     &&  &     &       &       \\

		    &
			EMSHS          & 9.69(0.25)    & 10.08(0.76)   & 0.51 & 0.78  & 105.68  
			&& 30.02(0.72)   & 30.13(2.21)   & 0.51 & 0.76 & 110.71  \\
		    $P1, V = 70,$ 
			&Blasso          & 15.73(0.71)   & 21.93(1.47)   & 0.51 & 0.78 & 7322.23 
			&& 46.72(2.00)   & 64.56(4.58)   & 0.51 & 0.55 & 6704.98 \\
			$\sigma^2 = 0.5$
			&SGL            & 71.82(4.08)   & 122.16(13.62) & 0.60 & 0.69  & 2201.40 
			&& 207.13(10.51) & 354.77(38.37) & 0.62 & 0.54 & 3557.34  \\
			&NRSS$_v$ &0.34(0.12)  &0.85(0.14)   &0.98  &1.00  & 1488.66 
			&& 23.03(0.90)  & 21.34(1.60)  &0.95   & 0.97 &  524.93\\
						&{\bf NRSS}        & 0.25(0.25)    & 0.84(0.36)    & 0.99 & 1.00 & 415.93  
			&& 23.51(0.98)   & 23.73(2.96)   & 0.97 & 0.97  & 455.47  \\

		\cmidrule{1-7} \cmidrule{9-13}\morecmidrules\cmidrule{1-7} \cmidrule{9-13}

			&
			EMSHS          & 7.27(0.21)   & 7.40(0.71)    & 0.51 & 0.68  & 16.88  
			&& 22.91(0.45)  & 22.70(1.71)   & 0.51 & 0.67 & 18.37   \\
			$P2, V = 20,$
			&Blasso          & 12.03(0.52)  & 16.88(1.46)   & 0.51 & 0.78	& 1641.95 
			&& 36.07(1.46)  & 49.05(4.45)   & 0.51 & 0.71 & 2468.57 \\
			$\sigma^2 = 0.1$
			&SGL            & 48.66(3.10)  & 87.93(10.34)  & 0.62 & 0.84 & 539.51  
			&& 143.66(7.71) & 245.90(26.38)  & 0.63 & 0.66  & 877.30  \\
			&NRSS$_v$  & 2.10(1.41) & 2.22(1.47)  & 0.88 &0.97  &362.35  
			&& 14.78(0.77)  & 13.38(1.2)  & 0.88  & 0.92 & 24.58 \\
						&{\bf NRSS}        & 0.31(0.06)   & 0.48(0.06)    & 0.98 & 1.00 & 252.18   
			&& 18.71(1.44)  & 19.19(2.55)   & 0.91 & 0.90     & 271.77  \\
			&            	&  &     &       &       &&  &     &       &       \\

		    &
			EMSHS          & 7.35(0.20)   & 7.88(0.73)    & 0.51 & 0.66  & 16.69  
			&& 22.99(0.44)  & 23.17(1.69)   & 0.51 & 0.67 & 17.30   \\
			$P2, V = 20,$
			&Blasso          & 12.09(0.58)  & 17.44(1.49)   & 0.51 & 0.76 & 1719.40   
			&& 36.29(1.63)  & 49.39(4.43)   & 0.51 & 0.70 & 1741.52 \\
			$\sigma^2 = 0.5$
			&SGL            & 50.18(2.85)  & 89.36(9.92)   & 0.60 & 0.79 & 630.52  
			&& 144.70(7.43) & 245.89(28.06) & 0.62 & 0.64  & 1009.81 \\
			&NRSS$_v$ & 3.92(1.18) &4.34(1.31)   &0.87  & 0.96 &511.30   
			&& 15.15(0.75)  & 14.14(1.25)  &0.88   & 0.92 &29.84 \\
						&{\bf NRSS}        & 2.61(0.60)   & 3.96(1.32)    & 0.93 & 0.99 & 318.35 
			&& 19.90(1.28)  & 21.21(2.38)   & 0.91 & 0.90  & 264.42  \\
			&            	&  &     &       &       &&  &     &       &       \\
			
		    &
			EMSHS          & 9.00(0.59)   & 9.19(0.94)     & 0.51 & 0.74  & 90.53  
			&& 28.70(0.43)  & 28.76(1.67)    & 0.51 & 0.71  & 88.44   \\
			$P2, V = 70,$
			&Blasso          & 14.61(0.48)  & 20.51(1.16)    & 0.51 & 0.76  & 6858.35
			&& 44.78(1.63)  & 62.13(3.40)    & 0.51 & 0.55 & 7095.14 \\
			$\sigma^2 = 0.1$
			&SGL            & 57.23(2.92)  & 97.98(10.18)   & 0.60 & 0.68 & 2498.82
			&& 175.00(7.49) & 294.12(26.03) & 0.61 & 0.52   & 4158.63 \\
			&NRSS$_v$  &0.38(0.42)  &0.45(0.22)   &0.98  &1.00  & 801.79 
			&&  23.13(0.80) &20.56(1.35)   &0.92   & 0.93 &  1271.02\\
						&{\bf NRSS}        & 0.13(0.03)   & 0.27(0.04)     & 0.98 & 1.00  & 599.60  
			&& 25.19(0.77)  & 25.57(2.40)    & 0.94 & 0.93  & 511.28  \\
			&            	&  &     &       &      &&  &     &       &       \\

			&
			EMSHS          & 9.02(0.17)   & 9.59(0.64)    & 0.51 & 0.73  & 77.63
			&& 28.76(0.42)  & 29.21(1.67)   & 0.51 & 0.71 & 84.23\\
		    $P2, V = 70,$
			&Blasso          & 14.74(0.55)  & 21.15(1.27)   & 0.51 & 0.74  & 6685.89
			&& 44.89(1.59)  & 62.84(3.57)   & 0.51 & 0.54   & 6779.87\\
		    $\sigma^2 = 0.5$
			&SGL            & 59.54(2.79)  & 101.42(9.72)  & 0.59 & 0.62  & 2620.43 
			&& 177.70(7.40) & 297.98(26.18) & 0.60 & 0.52 & 4650.01\\
			&NRSS$_v$ &	0.90(0.15)  &1.40(0.15)   & 0.96 &1.00  & 1125.81 
			&& 23.45(0.80)  &21.3(1.27)   &0.92   &0.92  & 	1215.79 \\
						&{\bf NRSS}        & 1.28(0.61)   & 2.37(0.69)    & 0.94 & 1.00  & 839.05
			&& 25.72(0.80)  & 26.81(2.48)   & 0.94 & 0.92  & 543.42\\
			\bottomrule 
	\end{tabular} }
\end{table}
    
Finally, from the computational perspective, NRSS achieves an efficient computation to handle such a complex brain connectome regression under  large-scale genetic variants. {When the prior connectivity is mis-specified as implemented by NRSS$_v$, we observe an increase in computational time.} Though the computational cost for NRSS is higher than that of EMSHS (which is also based on an EM algorithm but operates on a univariate outcome), NRSS displays 5- to 20-fold less computation cost compared with Blasso implemented by Gibbs sampler and the frequentist SGL.

{We now proceed to Simulation 2, where we set $P = 130,452$, $V = 87$, and $N = 1010$ to mimic the real data application in the presence of ultra-high dimensionality. We directly utilize the genetic data from the HCP as our $\bfx_i$. According to our results in Section \ref{sec:realdataHCP-YA}, we extract around 700 significant genetic signals to inform the locations in $\mathcal{B}$ where $\mathcal{B}_{:,:,p}$ has  non-zero elements  and  let $\mathcal{B}_{:,:,p}=\bfu_p \otimes \bfu_p$. The number of non-zero elements in $\bfu_p$ follows a Poisson distribution with a mean 8, and each non-zero element is simulated from $\mbox{N}(0, 0.1)$. We set random error $\sigma = 0.07$ yielding a SNR around 20. Following the same modeling procedure in Section \ref{sec:realdataHCP-YA}, we randomly split the data into train, test, and validation sets with sample sizes 506, 252, and 252 respectively and conduct a similar greedy search process to narrow the hyper-parameter combinations down to 12, under which 40 random splits are performed for each combination. We summarize the final selected features using stability selection by assigning $\pi_{thr}$ to have an upper bound of FDR around $0.3$ for SNPs according to \cite{meinshausen2010stability}. As for the competing methods, given their prohibitive computation under each ultra-high dimensional scenario, we only implement EMSHS ($\sim$2-5 days to train one EMSHS model using one Intel 2.50 Ghz core) on the data. For fair comparison, a similar model fitting procedure and result summary are applied.

We display the final stability selection results  in Table \ref{table:confusion} and estimation performance measured with $\text{MSE}_{\mathcal{B}}$ in Figure \ref{figure:MSE}. Clearly, the proposed NRSS achieves a much better selection accuracy with a sensitivity 0.61 and a specificity 1.00, compared with 0.42 and 0.98 obtained by EMSHS. The precision for NRSS is also higher than that for EMSHS (0.48 v.s. 0.13). We could draw a similar conclusion for the estimation performance with NRSS achieving a much smaller MSE for estimating  $\mathcal{B}$ as shown in Figure \ref{figure:MSE}. {To investigate the performance of type-I error controlling, we also present the  stability selection probabilities of top 400 selected SNPs under NRSS and EMSHS with true and false positives distinguished by different colors in Supplementary Figure 6. The results demonstrate that NRSS achieves satisfactory performance on FDR and type-I error under large thresholds of $\pi_{thr}$, and substantially outperforms the competing EMSHS.} Overall, NRSS maintains its  satisfactory performance under such ultra-high dimensional data. Together with Simulation 1, the computational efficiency and superior modeling performance of NRSS support its practical use in real-world brain imaging genetics applications.    }

\begin{minipage}{\textwidth}
 \begin{minipage}[b]{0.49\textwidth}
    \centering
    \scalebox{0.85}{
\begin{tabular}{@{}crrrcrr@{}}\toprule
& & \multicolumn{5}{c}{\bf Predicted}\\
& & \multicolumn{2}{c}{NRSS} & \phantom{abc}& \multicolumn{2}{c}{EMSHS} \\
\cmidrule{3-4} \cmidrule{6-7}
 & &  Yes & No && Yes & No\\ \midrule
\multirow{ 2}{*}{{\bf Actual} }&
Yes & 432 & 282 && 300 & 414\\
& No &  460 & 129278 && 2034 & 127704\\
\bottomrule
\end{tabular}}
\captionof{table}{\centering    Confusion matrices for NRSS and EMSHS under Simulation 2. \label{table:confusion}}
    \end{minipage}
  \begin{minipage}[b]{0.49\textwidth}
    \centering
     \includegraphics[width=6.7cm]{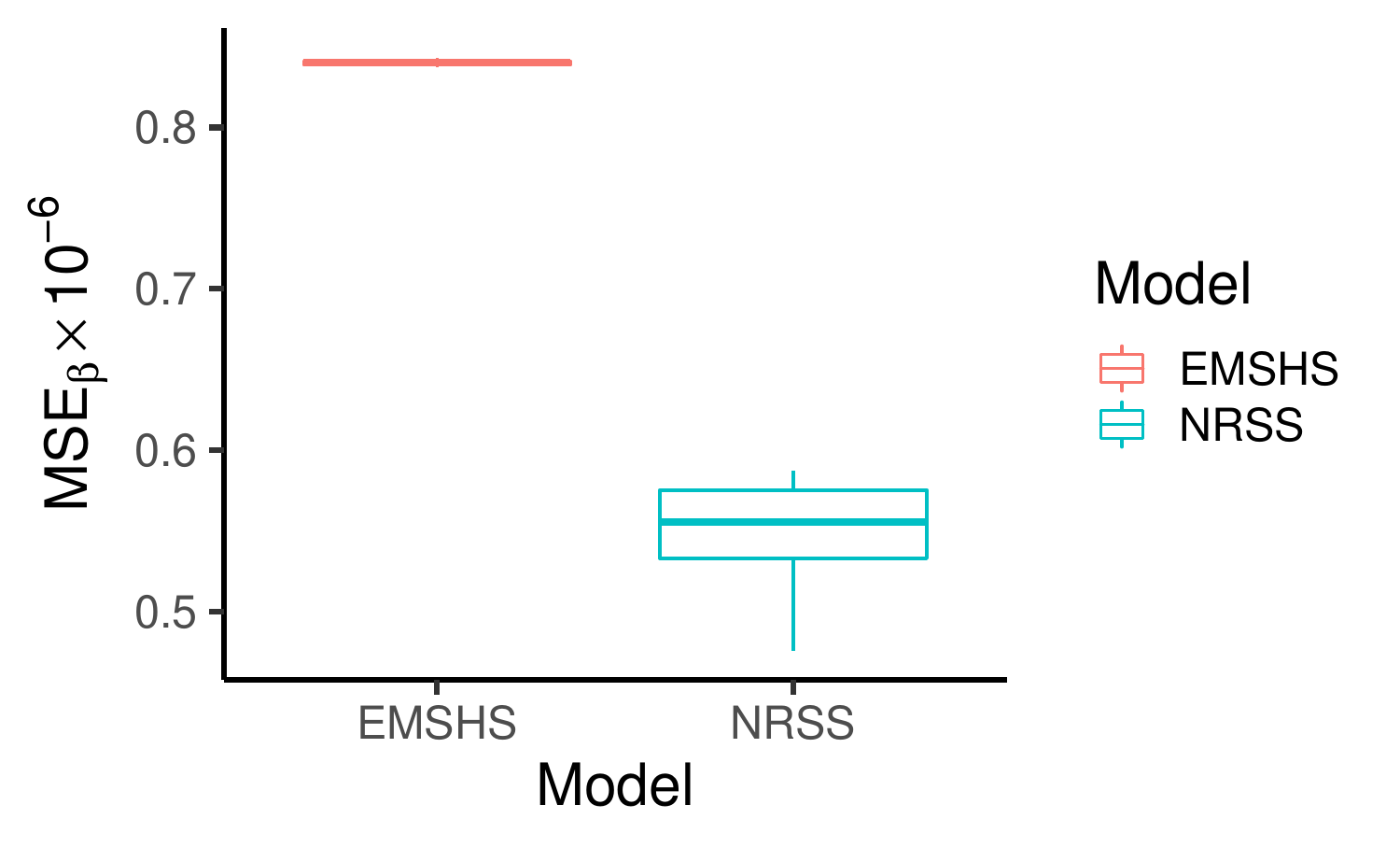}\vspace{-0.1cm}
    \captionof{figure}{\centering Box plot of $\text{MSE}_{\mathcal{B}}$ for NRSS and EMSHS under Simulation 2. \label{figure:MSE}}
  \end{minipage}
  \hfill

  \end{minipage}
  \vspace{0.1cm}

\section{Conclusion}\label{sec:dicussion}
Motivated by the need to understand how genetic factors influence human brain structural connectome throughout young adulthood, we develop a generalized Bayesian regression model for a network response under high dimensional structured predictors to identify genetic signals and their corresponding brain white matter fiber sub-networks. The proposed method maintains the topological structure for a brain connectome by modeling clique sub-graphs and enhances the analytical power and biological interpretability through an integrative structural shrinkage construction. To alleviate computational challenges, we develop an efficient EM algorithm to perform posterior inference. We show the superiority of our model in extensive simulations. In the application to the HCP-YA dataset, we discover highly interpretable genetic underpinnings including the selected biomarkers in the synaptic vesicle cycle pathway which are highly associated with the cognitive and memory related cross-hemispheric and cortical to sub-cortical fiber connections.

We focus on modeling the brain structural connectome which is sparse in nature. When it comes to other brain connectomes, say the functional connectome which is much more dense, we could modify model \eqref{eqn:model2} through a different low-rank representation, e.g., using stochastic block models to explicitly capture the sub-network module blocks.  In general, the parameterization for the response network in  \eqref{eqn:model} can be flexibly tailored for different application needs. For the genetic component, we propose here to study individual variant's impact and aggregate signal via the grouping effect along SNPs. Alternatively, we could also modify model  \eqref{eqn:model} to a network response mixed effect model to quantify additive SNP heritability for the structural connectome. 
Finally, we target a healthy sub-population to investigate normal brain variations impacted by genetics. Given the emerging trend to study how genetic risk variants and environmental interventions contribute to brain structural abnormalities, we plan to further incorporate additional data on environmental factors to study this integrative effect pathways between genetics, environment and brain connectome under vulnerable sub-populations.

{Finally, we apply the proposed method to the whole samples in the HCP-YA dataset acknowledging the existence of twins/siblings in the study. There is a broad literature on whether and how sample relatedness should be accommodated in genetic association analyses \citep{nazarian2020impact}. In general, for quantitative traits including the brain structural network, the issue of correctly uncovering the variance component under correlated samples is not as severe as that for binary traits \citep{visscher2008genome}. Without pursuing a traditional inference procedure as in our case with an MAP estimator, we anticipate that the impact of this 
correlated structure on our results, especially the predictive performance, is limited. Though we develop the current model as an initial effort for understanding the relationship between genetics and the brain structural network, we plan to incorporate pedigree information in the model as a future extension.    }

\section*{Supplementary Materials}
Please email the corresponding authors for the supplementary materials, which contain additional figures, ROI information and results for data application and simulations, and the programming code to implement the method and perform all the numerical studies. 




\bibliographystyle{asa}
\baselineskip=10pt
\bibliography{paperdti}

\begin{thebibliography}{66}
\newcommand{\enquote}[1]{``#1''}
\expandafter\ifx\csname natexlab\endcsname\relax\def\natexlab#1{#1}\fi

\bibitem[{Alba(1973)}]{alba1973graph}
Alba, R.~D. (1973), \enquote{A graph-theoretic definition of a sociometric
  clique,} \textit{Journal of Mathematical Sociology}, 3, 113--126.

\bibitem[{Antonenko et~al.(2013)Antonenko, Brauer, Meinzer, Fengler, Kerti,
  Friederici, and Fl{\"o}el}]{antonenko2013functional}
Antonenko, D., Brauer, J., Meinzer, M., Fengler, A., Kerti, L., Friederici,
  A.~D., and Fl{\"o}el, A. (2013), \enquote{Functional and structural syntax
  networks in aging,} \textit{Neuroimage}, 83, 513--523.

\bibitem[{Arnatkeviciute et~al.(2021)Arnatkeviciute, Fulcher, Bellgrove, and
  Fornito}]{arnatkeviciute2021genome}
Arnatkeviciute, A., Fulcher, B., Bellgrove, M., and Fornito, A. (2021),
  \enquote{Where the genome meets the connectome: understanding how genes shape
  human brain connectivity,} \textit{NeuroImage}, 244, 118570.

\bibitem[{Bassett and Sporns(2017)}]{bassett2017network}
Bassett, D.~S. and Sporns, O. (2017), \enquote{Network neuroscience,}
  \textit{Nature Neuroscience}, 20, 353--364.

\bibitem[{Beck and Teboulle(2009)}]{Beck2009}
Beck, A. and Teboulle, M. (2009), \enquote{A Fast Iterative
  Shrinkage-Thresholding Algorithm for Linear Inverse Problems,} \textit{SIAM
  Journal on Imaging Sciences}, 2, 183--202.

\bibitem[{Butt et~al.(2014)Butt, Fern, and Matute}]{butt2014neurotransmitter}
Butt, A.~M., Fern, R.~F., and Matute, C. (2014), \enquote{Neurotransmitter
  signaling in white matter,} \textit{Glia}, 62, 1762--1779.

\bibitem[{Carvalho et~al.(2010)Carvalho, Polson, and
  Scott}]{carvalho2010horseshoe}
Carvalho, C.~M., Polson, N.~G., and Scott, J.~G. (2010), \enquote{The horseshoe
  estimator for sparse signals,} \textit{Biometrika}, 97, 465--480.

\bibitem[{Chang et~al.(2018)Chang, Kundu, and Long}]{chang2018}
Chang, C., Kundu, S., and Long, Q. (2018), \enquote{Scalable Bayesian variable
  selection for structured high-dimensional data,} \textit{Biometrics}, 74,
  1372--1382.

\bibitem[{Chiang et~al.(2011)Chiang, McMahon, de~Zubicaray, Martin, Hickie,
  Toga, Wright, and Thompson}]{chiang2011genetics}
Chiang, M.-C., McMahon, K.~L., de~Zubicaray, G.~I., Martin, N.~G., Hickie, I.,
  Toga, A.~W., Wright, M.~J., and Thompson, P.~M. (2011), \enquote{Genetics of
  white matter development: a DTI study of 705 twins and their siblings aged 12
  to 29,} \textit{NeuroImage}, 54, 2308--2317.

\bibitem[{Choi et~al.(2016)Choi, Han, Kang, Won, Chang, Tae, Son, Kim, Lee, and
  Ham}]{choi2016effects}
Choi, S., Han, K.-M., Kang, J., Won, E., Chang, H.~S., Tae, W.~S., Son, K.~R.,
  Kim, S.-J., Lee, M.-S., and Ham, B.-J. (2016), \enquote{Effects of a
  polymorphism of the neuronal amino acid transporter SLC6A15 gene on
  structural integrity of white matter tracts in major depressive disorder,}
  \textit{PloS one}, 11, e0164301.

\bibitem[{Consortium et~al.(2010)}]{international2010integrating}
Consortium, I. H.~. et~al. (2010), \enquote{Integrating common and rare genetic
  variation in diverse human populations,} \textit{Nature}, 467, 52.

\bibitem[{Desikan et~al.(2006)Desikan, Ségonne, Fischl, Quinn, Dickerson,
  Blacker, Buckner, Dale, Maguire, Hyman, Albert, and
  Killiany}]{Desikan2006968}
Desikan, R.~S., Ségonne, F., Fischl, B., Quinn, B.~T., Dickerson, B.~C.,
  Blacker, D., Buckner, R.~L., Dale, A.~M., Maguire, R.~P., Hyman, B.~T.,
  Albert, M.~S., and Killiany, R.~J. (2006), \enquote{An automated labeling
  system for subdividing the human cerebral cortex on {MRI} scans into gyral
  based regions of interest,} \textit{NeuroImage}, 31, 968 -- 980.

\bibitem[{Durante et~al.(2018)Durante, Dunson, et~al.}]{durante2018bayesian}
Durante, D., Dunson, D.~B., et~al. (2018), \enquote{Bayesian inference and
  testing of group differences in brain networks,} \textit{Bayesian Analysis},
  13, 29--58.

\bibitem[{Elliott et~al.(2018)Elliott, Sharp, Alfaro-Almagro, Shi, Miller,
  Douaud, Marchini, and Smith}]{elliott2018genome}
Elliott, L.~T., Sharp, K., Alfaro-Almagro, F., Shi, S., Miller, K.~L., Douaud,
  G., Marchini, J., and Smith, S.~M. (2018), \enquote{Genome-wide association
  studies of brain imaging phenotypes in UK Biobank,} \textit{Nature}, 562,
  210--216.

\bibitem[{Elsheikh et~al.(2020)Elsheikh, Chimusa, Mulder, and
  Crimi}]{elsheikh2020genome}
Elsheikh, S.~S., Chimusa, E.~R., Mulder, N.~J., and Crimi, A. (2020),
  \enquote{Genome-wide association study of brain connectivity changes for
  alzheimer’s disease,} \textit{Scientific Reports}, 10, 1--16.

\bibitem[{Giddaluru et~al.(2016)Giddaluru, Espeseth, Salami, Westlye,
  Lundquist, Christoforou, Cichon, Adolfsson, Steen, Reinvang,
  et~al.}]{giddaluru2016genetics}
Giddaluru, S., Espeseth, T., Salami, A., Westlye, L.~T., Lundquist, A.,
  Christoforou, A., Cichon, S., Adolfsson, R., Steen, V.~M., Reinvang, I.,
  et~al. (2016), \enquote{Genetics of structural connectivity and information
  processing in the brain,} \textit{Brain Structure and Function}, 221,
  4643--4661.

\bibitem[{Hoff et~al.(2002)Hoff, Raftery, and Handcock}]{hoff2002latent}
Hoff, P.~D., Raftery, A.~E., and Handcock, M.~S. (2002), \enquote{Latent space
  approaches to social network analysis,} \textit{Journal of the American
  Statistical Association}, 97, 1090--1098.

\bibitem[{Hu et~al.(2020)Hu, Kong, and Shen}]{hu2019nonparametric}
Hu, W., Kong, D., and Shen, W. (2020), \enquote{Nonparametric Matrix Response
  Regression with Application to Brain Imaging Data Analysis,}
  \textit{Biometrics}, https://doi.org/10.1111/biom.13362.

\bibitem[{Hulshagen et~al.(2008)Hulshagen, Krysko, Bottelbergs, Huyghe, Klein,
  Van~Veldhoven, De~Deyn, D'Hooge, Hartmann, and Baes}]{hulshagen2008absence}
Hulshagen, L., Krysko, O., Bottelbergs, A., Huyghe, S., Klein, R.,
  Van~Veldhoven, P.~P., De~Deyn, P.~P., D'Hooge, R., Hartmann, D., and Baes, M.
  (2008), \enquote{Absence of functional peroxisomes from mouse CNS causes
  dysmyelination and axon degeneration,} \textit{Journal of Neuroscience}, 28,
  4015--4027.

\bibitem[{Kassmann(2014)}]{kassmann2014myelin}
Kassmann, C.~M. (2014), \enquote{Myelin peroxisomes--Essential organelles for
  the maintenance of white matter in the nervous system,} \textit{Biochimie},
  98, 111--118.

\bibitem[{Kennedy(2016)}]{kennedy2016synaptic}
Kennedy, M.~B. (2016), \enquote{Synaptic signaling in learning and memory,}
  \textit{Cold Spring Harbor Perspectives in Biology}, 8, a016824.

\bibitem[{Knierim(2015)}]{knierim2015hippocampus}
Knierim, J.~J. (2015), \enquote{The hippocampus,} \textit{Current Biology}, 25,
  R1116--R1121.

\bibitem[{Kong et~al.(2020)Kong, An, Zhang, and Zhu}]{kong2020l2rm}
Kong, D., An, B., Zhang, J., and Zhu, H. (2020), \enquote{L2RM: Low-Rank Linear
  Regression Models for High-Dimensional Matrix Responses,} \textit{Journal of
  the American Statistical Association}, 115, 403--424.

\bibitem[{Lenroot and Giedd(2008)}]{lenroot2008changing}
Lenroot, R.~K. and Giedd, J.~N. (2008), \enquote{The changing impact of genes
  and environment on brain development during childhood and adolescence:
  initial findings from a neuroimaging study of pediatric twins,}
  \textit{Development and Psychopathology}, 20, 1161.

\bibitem[{Li et~al.(2021)Li, Hu, Wang, and Zhang}]{li2020super}
Li, T., Hu, J., Wang, S., and Zhang, H. (2021), \enquote{Super-variants
  identification for brain connectivity,} \textit{Human Brain Mapping}, 42,
  1304--1312.

\bibitem[{Lisman et~al.(2017)Lisman, Buzs{\'a}ki, Eichenbaum, Nadel, Ranganath,
  and Redish}]{lisman2017viewpoints}
Lisman, J., Buzs{\'a}ki, G., Eichenbaum, H., Nadel, L., Ranganath, C., and
  Redish, A.~D. (2017), \enquote{Viewpoints: how the hippocampus contributes to
  memory, navigation and cognition,} \textit{Nature Neuroscience}, 20,
  1434--1447.

\bibitem[{Liu and Xie(2020)}]{liu2020cauchy}
Liu, Y. and Xie, J. (2020), \enquote{Cauchy combination test: a powerful test
  with analytic p-value calculation under arbitrary dependency structures,}
  \textit{Journal of the American Statistical Association}, 115, 393--402.

\bibitem[{Maier-Hein et~al.(2017)Maier-Hein, Neher, Houde, C{\^o}t{\'e},
  Garyfallidis, Zhong, Chamberland, Yeh, Lin, Ji,
  et~al.}]{maier2016tractography}
Maier-Hein, K.~H., Neher, P.~F., Houde, J.-C., C{\^o}t{\'e}, M.-A.,
  Garyfallidis, E., Zhong, J., Chamberland, M., Yeh, F.-C., Lin, Y.-C., Ji, Q.,
  et~al. (2017), \enquote{The challenge of mapping the human connectome based
  on diffusion tractography,} \textit{Nature Communications}, 8, 1349.

\bibitem[{Meinshausen and B{\"u}hlmann(2010)}]{meinshausen2010stability}
Meinshausen, N. and B{\"u}hlmann, P. (2010), \enquote{Stability selection,}
  \textit{Journal of the Royal Statistical Society: Series B (Statistical
  Methodology)}, 72, 417--473.

\bibitem[{Moser and Moser(1998)}]{moser1998functional}
Moser, M.-B. and Moser, E.~I. (1998), \enquote{Functional differentiation in
  the hippocampus,} \textit{Hippocampus}, 8, 608--619.

\bibitem[{Nathoo et~al.(2019)Nathoo, Kong, Zhu, and
  Initiative}]{nathoo2019review}
Nathoo, F.~S., Kong, L., Zhu, H., and Initiative, A. D.~N. (2019), \enquote{A
  review of statistical methods in imaging genetics,} \textit{Canadian Journal
  of Statistics}, 47, 108--131.

\bibitem[{Nazarian et~al.(2020)Nazarian, Arbeev, and
  Kulminski}]{nazarian2020impact}
Nazarian, A., Arbeev, K.~G., and Kulminski, A.~M. (2020), \enquote{The impact
  of disregarding family structure on genome-wide association analysis of
  complex diseases in cohorts with simple pedigrees,} \textit{Journal of
  applied genetics}, 61, 75--86.

\bibitem[{Park and Friston(2013)}]{park2013structural}
Park, H.-J. and Friston, K. (2013), \enquote{Structural and functional brain
  networks: from connections to cognition,} \textit{Science}, 342, 1238411.

\bibitem[{Park and Casella(2008)}]{park2008bayesian}
Park, T. and Casella, G. (2008), \enquote{The bayesian lasso,} \textit{Journal
  of the American Statistical Association}, 103, 681--686.

\bibitem[{Powell(2006)}]{powell2006gene}
Powell, C.~M. (2006), \enquote{Gene targeting of presynaptic proteins in
  synaptic plasticity and memory: across the great divide,}
  \textit{Neurobiology of Learning and Memory}, 85, 2--15.

\bibitem[{Ramanan et~al.(2014)Ramanan, Risacher, Nho, Kim, Swaminathan, Shen,
  Foroud, Hakonarson, Huentelman, Aisen, et~al.}]{ramanan2014apoe}
Ramanan, V.~K., Risacher, S.~L., Nho, K., Kim, S., Swaminathan, S., Shen, L.,
  Foroud, T.~M., Hakonarson, H., Huentelman, M.~J., Aisen, P.~S., et~al.
  (2014), \enquote{APOE and BCHE as modulators of cerebral amyloid deposition:
  a florbetapir PET genome-wide association study,} \textit{Molecular
  Psychiatry}, 19, 351--357.

\bibitem[{Ramasamy et~al.(2014)Ramasamy, Trabzuni, Guelfi, Varghese, Smith,
  Walker, De, Hardy, Ryten, Weale, et~al.}]{ramasamy2014genetic}
Ramasamy, A., Trabzuni, D., Guelfi, S., Varghese, V., Smith, C., Walker, R.,
  De, T., Hardy, J., Ryten, M., Weale, M.~E., et~al. (2014), \enquote{Genetic
  variability in the regulation of gene expression in ten regions of the human
  brain,} \textit{Nature Neuroscience}, 17, 1418--1428.

\bibitem[{Reli{\'o}n et~al.(2019)Reli{\'o}n, Kessler, Levina, Taylor,
  et~al.}]{relion2019network}
Reli{\'o}n, J. D.~A., Kessler, D., Levina, E., Taylor, S.~F., et~al. (2019),
  \enquote{Network classification with applications to brain connectomics,}
  \textit{The Annals of Applied Statistics}, 13, 1648--1677.

\bibitem[{Richiardi et~al.(2015)Richiardi, Altmann, Milazzo, Chang,
  Chakravarty, Banaschewski, Barker, Bokde, Bromberg, B{\"u}chel,
  et~al.}]{richiardi2015correlated}
Richiardi, J., Altmann, A., Milazzo, A.-C., Chang, C., Chakravarty, M.~M.,
  Banaschewski, T., Barker, G.~J., Bokde, A.~L., Bromberg, U., B{\"u}chel, C.,
  et~al. (2015), \enquote{Correlated gene expression supports synchronous
  activity in brain networks,} \textit{Science}, 348, 1241--1244.

\bibitem[{Satizabal et~al.(2019)Satizabal, Adams, Hibar, White, Knol, Stein,
  Scholz, Sargurupremraj, Jahanshad, Roshchupkin,
  et~al.}]{satizabal2019genetic}
Satizabal, C.~L., Adams, H.~H., Hibar, D.~P., White, C.~C., Knol, M.~J., Stein,
  J.~L., Scholz, M., Sargurupremraj, M., Jahanshad, N., Roshchupkin, G.~V.,
  et~al. (2019), \enquote{Genetic architecture of subcortical brain structures
  in 38,851 individuals,} \textit{Nature Genetics}, 51, 1624--1636.

\bibitem[{Shen and Thompson(2019)}]{shen2019brain}
Shen, L. and Thompson, P.~M. (2019), \enquote{Brain imaging genomics:
  integrated analysis and machine learning,} \textit{Proceedings of the IEEE},
  108, 125--162.

\bibitem[{Shen et~al.(2017)Shen, Finn, Scheinost, Rosenberg, Chun,
  Papademetris, and Constable}]{shen2017using}
Shen, X., Finn, E.~S., Scheinost, D., Rosenberg, M.~D., Chun, M.~M.,
  Papademetris, X., and Constable, R.~T. (2017), \enquote{Using
  connectome-based predictive modeling to predict individual behavior from
  brain connectivity,} \textit{Nature Protocols}, 12, 506--518.

\bibitem[{S{\"u}dhof(1995)}]{sudhof1995synaptic}
S{\"u}dhof, T.~C. (1995), \enquote{The synaptic vesicle cycle: a cascade of
  protein--protein interactions,} \textit{Nature}, 375, 645--653.

\bibitem[{Sun and Li(2017)}]{sun2017store}
Sun, W.~W. and Li, L. (2017), \enquote{STORE: sparse tensor response regression
  and neuroimaging analysis,} \textit{The Journal of Machine Learning
  Research}, 18, 4908--4944.

\bibitem[{Tao et~al.(2017)Tao, Nichols, Hua, Ching, Rolls, Thompson, Feng,
  Initiative, et~al.}]{tao2017generalized}
Tao, C., Nichols, T.~E., Hua, X., Ching, C.~R., Rolls, E.~T., Thompson, P.~M.,
  Feng, J., Initiative, A. D.~N., et~al. (2017), \enquote{Generalized reduced
  rank latent factor regression for high dimensional tensor fields, and
  neuroimaging-genetic applications,} \textit{NeuroImage}, 144, 35--57.

\bibitem[{Van~Essen et~al.(2013)Van~Essen, Smith, Barch, Behrens, Yacoub,
  Ugurbil, Consortium, et~al.}]{van2013wu}
Van~Essen, D.~C., Smith, S.~M., Barch, D.~M., Behrens, T.~E., Yacoub, E.,
  Ugurbil, K., Consortium, W.-M.~H., et~al. (2013), \enquote{The {WU-Minn}
  human connectome project: an overview,} \textit{NeuroImage}, 80, 62--79.

\bibitem[{Van~Essen et~al.(2012)Van~Essen, Ugurbil, Auerbach, Barch, Behrens,
  Bucholz, Chang, Chen, Corbetta, Curtiss, et~al.}]{VanEssen20122222}
Van~Essen, D.~C., Ugurbil, K., Auerbach, E., Barch, D., Behrens, T., Bucholz,
  R., Chang, A., Chen, L., Corbetta, M., Curtiss, S.~W., et~al. (2012),
  \enquote{The Human Connectome Project: a data acquisition perspective,}
  \textit{NeuroImage}, 62, 2222--2231.

\bibitem[{Villabona-Rueda et~al.(2019)Villabona-Rueda, Erice, Pardo, and
  Stins}]{villabona2019evolving}
Villabona-Rueda, A., Erice, C., Pardo, C.~A., and Stins, M.~F. (2019),
  \enquote{The evolving concept of the blood brain barrier (BBB): from a single
  static barrier to a heterogeneous and dynamic relay center,}
  \textit{Frontiers in Cellular Neuroscience}, 13, 405.

\bibitem[{Visscher et~al.(2008)Visscher, Andrew, and
  Nyholt}]{visscher2008genome}
Visscher, P.~M., Andrew, T., and Nyholt, D.~R. (2008), \enquote{Genome-wide
  association studies of quantitative traits with related individuals: little
  (power) lost but much to be gained,} \textit{European Journal of Human
  Genetics}, 16, 387--390.

\bibitem[{Wang et~al.(2021)Wang, Lin, Cole, and Zhang}]{WANG2021117493}
Wang, L., Lin, F.~V., Cole, M., and Zhang, Z. (2021), \enquote{Learning Clique
  Subgraphs in Structural Brain Network Classification with Application to
  Crystallized Cognition,} \textit{NeuroImage}, 225, 117493.

\bibitem[{Wang and Zhang(2019)}]{wang2019learning}
Wang, L. and Zhang, Z. (2019), \enquote{Learning Signal Subgraphs from
  Longitudinal Brain Networks with Symmetric Bilinear Logistic Regression,}
  \textit{arXiv preprint arXiv:1908.05627}.

\bibitem[{Wang et~al.(2019)Wang, Zhang, Dunson, et~al.}]{wang2019common}
Wang, L., Zhang, Z., Dunson, D., et~al. (2019), \enquote{Common and individual
  structure of brain networks,} \textit{The Annals of Applied Statistics}, 13,
  85--112.

\bibitem[{Webster and Descoteaux(2015)}]{webster2015high}
Webster, J. and Descoteaux, M. (2015), \enquote{High Angular Resolution
  Diffusion Imaging (HARDI),} \textit{Wiley Encyclopedia of Electrical and
  Electronics Engineering}.

\bibitem[{Wu et~al.(2010)Wu, Kraft, Epstein, Taylor, Chanock, Hunter, and
  Lin}]{wu2010powerful}
Wu, M.~C., Kraft, P., Epstein, M.~P., Taylor, D.~M., Chanock, S.~J., Hunter,
  D.~J., and Lin, X. (2010), \enquote{Powerful SNP-set analysis for
  case-control genome-wide association studies,} \textit{The American Journal
  of Human Genetics}, 86, 929--942.

\bibitem[{Wu et~al.(2011)Wu, Lee, Cai, Li, Boehnke, and Lin}]{wu2011rare}
Wu, M.~C., Lee, S., Cai, T., Li, Y., Boehnke, M., and Lin, X. (2011),
  \enquote{Rare-variant association testing for sequencing data with the
  sequence kernel association test,} \textit{The American Journal of Human
  Genetics}, 89, 82--93.

\bibitem[{Yao et~al.(2020)Yao, Cong, Yan, Risacher, Saykin, Moore, Shen,
  Consortium, and Initiative}]{yao2020regional}
Yao, X., Cong, S., Yan, J., Risacher, S.~L., Saykin, A.~J., Moore, J.~H., Shen,
  L., Consortium, U. B.~E., and Initiative, A. D.~N. (2020), \enquote{Regional
  imaging genetic enrichment analysis,} \textit{Bioinformatics}, 36,
  2554--2560.

\bibitem[{Yuan and Lin(2006)}]{yuan2006model}
Yuan, M. and Lin, Y. (2006), \enquote{Model selection and estimation in
  regression with grouped variables,} \textit{Journal of the Royal Statistical
  Society: Series B (Statistical Methodology)}, 68, 49--67.

\bibitem[{Zhang et~al.(2016)Zhang, Cheng, Liu, Zhang, Lei, Yao, Becker, Liu,
  Kendrick, Lu, et~al.}]{zhang2016neural}
Zhang, J., Cheng, W., Liu, Z., Zhang, K., Lei, X., Yao, Y., Becker, B., Liu,
  Y., Kendrick, K.~M., Lu, G., et~al. (2016), \enquote{Neural,
  electrophysiological and anatomical basis of brain-network variability and
  its characteristic changes in mental disorders,} \textit{Brain}, 139,
  2307--2321.

\bibitem[{Zhang et~al.(2018{\natexlab{a}})Zhang, Sun, and
  Li}]{zhang2018network}
Zhang, J., Sun, W.~W., and Li, L. (2018{\natexlab{a}}), \enquote{Network
  Response Regression for Modeling Population of Networks with Covariates,}
  \textit{arXiv preprint arXiv:1810.03192}.

\bibitem[{Zhang et~al.(2019)Zhang, Allen, Zhu, and Dunson}]{zhang2019tensor}
Zhang, Z., Allen, G.~I., Zhu, H., and Dunson, D. (2019), \enquote{Tensor
  network factorizations: Relationships between brain structural connectomes
  and traits,} \textit{NeuroImage}, 197, 330--343.

\bibitem[{Zhang et~al.(2018{\natexlab{b}})Zhang, Descoteaux, Zhang, Girard,
  Chamberland, Dunson, Srivastava, and Zhu}]{Zhang2017HCP}
Zhang, Z., Descoteaux, M., Zhang, J., Girard, G., Chamberland, M., Dunson, D.,
  Srivastava, A., and Zhu, H. (2018{\natexlab{b}}), \enquote{Mapping
  Population-based Structural Connectomes,} \textit{NeuroImage}, 172, 130 --
  145.

\bibitem[{Zhao et~al.(2021{\natexlab{a}})Zhao, Li, Yang, Wang, Luo, Shan, Zhu,
  Xiong, Hauberg, Bendl, et~al.}]{zhao2021common}
Zhao, B., Li, T., Yang, Y., Wang, X., Luo, T., Shan, Y., Zhu, Z., Xiong, D.,
  Hauberg, M.~E., Bendl, J., et~al. (2021{\natexlab{a}}), \enquote{Common
  genetic variation influencing human white matter microstructure,}
  \textit{Science}, 372.

\bibitem[{Zhao et~al.(2019{\natexlab{a}})Zhao, Zhang, Ibrahim, Luo, Santelli,
  Li, Li, Shan, Zhu, Zhou, et~al.}]{zhao2019large}
Zhao, B., Zhang, J., Ibrahim, J.~G., Luo, T., Santelli, R.~C., Li, Y., Li, T.,
  Shan, Y., Zhu, Z., Zhou, F., et~al. (2019{\natexlab{a}}),
  \enquote{Large-scale GWAS reveals genetic architecture of brain white matter
  microstructure and genetic overlap with cognitive and mental health traits
  (n= 17,706),} \textit{Molecular Psychiatry}, 1--13.

\bibitem[{Zhao et~al.(2015)Zhao, Kang, and Long}]{zhao2015bayesian}
Zhao, Y., Kang, J., and Long, Q. (2015), \enquote{Bayesian multiresolution
  variable selection for ultra-high dimensional neuroimaging data,}
  \textit{IEEE/ACM transactions on computational biology and bioinformatics},
  15, 537--550.

\bibitem[{Zhao et~al.(2021{\natexlab{b}})Zhao, Wang, Mostofsky, Caffo, and
  Luo}]{zhao2019covariate}
Zhao, Y., Wang, B., Mostofsky, S.~H., Caffo, B.~S., and Luo, X.
  (2021{\natexlab{b}}), \enquote{Covariate Assisted Principal regression for
  covariance matrix outcomes,} \textit{Biostatistics}, 22, 629--645.

\bibitem[{Zhao et~al.(2019{\natexlab{b}})Zhao, Zhu, Lu, Knickmeyer, and
  Zou}]{zhao2019structured}
Zhao, Y., Zhu, H., Lu, Z., Knickmeyer, R.~C., and Zou, F. (2019{\natexlab{b}}),
  \enquote{Structured Genome-Wide Association Studies with Bayesian
  Hierarchical Variable Selection,} \textit{Genetics}, 212, 397--415.

\end{thebibliography}

\end{document}